\documentclass[preprint,aps ,nofootinbib]{revtex4}
\pdfoutput=1
\usepackage[colorlinks=true,linkcolor=blue,urlcolor=blue,filecolor=black,citecolor=red,pdfstartview=FitV,pdftitle={},pdfsubject={},pdfkeywords={},pdfpagemode=None,bookmarksopen=true]{hyperref}
\usepackage{graphicx}
\usepackage{epsfig}
\usepackage{amsmath}
\usepackage{amsfonts}
\usepackage{amssymb}
\usepackage{color}%
\usepackage{slashed}
\usepackage{dcolumn}
\providecommand{\U}[1]{\protect\rule{.1in}{.1in}}

\newcommand{\f}{\begin{equation}}
\newcommand{\ff}{\end{equation}}
\newcommand{\fa}{\begin{eqnarray}}
\newcommand{\ffa}{\end{eqnarray}}

\allowdisplaybreaks[4]

\begin{document}
\title{Holographic fermionic spectrum from Born-Infeld AdS black hole}
\author{Jian-Pin Wu $^{1,2}$}
\email{jianpinwu@mail.bnu.edu.cn}
\affiliation{$^{1}$Institute of Gravitation and Cosmology, Department of Physics, School of Mathematics and Physics, Bohai University, Jinzhou 121013, China
\\$^{2}$Shanghai Key Laboratory of High Temperature Superconductors, Shanghai, 200444, China.}
\begin{abstract}

In this letter, we systematically explore the holographic (non-)relativistic fermionic spectrum without/with dipole coupling dual to Born-Infeld anti-de Sitter (BI-AdS) black hole.
For the relativistic fermionic fixed point, this holographic fermionic system exhibits non-Fermi liquid behavior.
Also, with the increase of BI parameter $\gamma$, the non-Fermi liquid becomes even ``more non-Fermi".
When the dipole coupling term is included, we find that the BI term makes it a lot tougher to form the gap.
While for the non-relativistic fermionic system with large dipole coupling in BI-AdS background,
with the increase of BI parameter, the gap comes into being against.

\end{abstract} \maketitle

\section{Introduction}

By now there are not still a well-established theoretical framework to describe and understand
the strange metal phase and the Mott gap which are usually attributed to the strongly correlated effects.
Recently AdS/CFT correspondence provides us with a new and operative approach to attack these problems.
The profound influential examples are the building of the holographic non-Fermi liquids \cite{Liu:2009dm},
the emergence of Mott gap \cite{Edalati:2010ww} and flat band \cite{Laia:2011zn} from holographic fermionic system.
In this letter, we shall further explore these problems in the Born-Infeld anti-de Sitter (BI-AdS) geometry.

In \cite{Liu:2009dm},
they explore the holographic fermionic response over the Reissner-Nordstr$\ddot{\texttt{o}}$m (RN-AdS) back hole.
In particular, they numerically study the scaling behavior near the Fermi surface and find that it exhibits a non-linear dispersion relation \cite{Liu:2009dm}.
It indicates that this holographic fermionic system can model non-Fermi liquid behavior which is an example of strangle metal phase.
Furthermore, they find that this scaling behavior is controlled by conformal dimensions in the IR CFT dual to AdS$_2$ \cite{Faulkner:2009wj}.
Subsequently, a lot of extensive explorations on the Fermi surface structure and associated excitations have also been implemented in
more general geometries in \cite{Wu:2011bx,Liu:2012tr,Ling:2013aya,Wu:2011cy,Li:2011sh,Gursoy:2011gz,Alishahiha:2012nm,
Fang:2012pw,Li:2012uua,Wang:2013tv,Wu:2013xta,Fang:2013ixa,Kuang:2014pna,Fang:2014jka,Fang:2015vpa,Fang:2015dia} and references therein.
These holographic fermionic systems are expected to be candidates for generalized non-Fermi liquids and offer a possible clue to uncover the basic principle hidden behind the strangle metal phase.

While the chiral symmetry-breaking dipole coupling term is introduced,
a Mott gap emerges in the fermionic spectral function \cite{Edalati:2010ww,Edalati:2010ge},
which indicates Mott phase is implemented in the holographic framework.
Besides the Mott hard gap, they also find that the spectral weight transfer between bands, which is one of the characteristic of doped Mott insulator.
Further, the fermionic spectrum in presence of dipole coupling term in other background have also been explored in \cite{Wu:2012fk,Kuang:2012ud,Wen:2012ur,Kuang:2012tq,Wu:2013oea,Wu:2014rqa,Kuang:2014yya,Ling:2014bda,Vanacore:2015,Fan:2013zqa}.
These studies further confirm that the emergence of Mott gap is robust when the dipole coupling term is introduced.

On the other hand, when Lorentz violating boundary terms are imposed on the Dirac spinor field,
a non-relativistic fermionic fixed point can be implemented in AdS/CFT \cite{Laia:2011wf}.
This dual boundary theory exhibits a dispersionless flat band \cite{Laia:2011zn}.
Its low energy behavior is also analytically explored in \cite{Wu:2013vma} and they
find that the scaling behavior is also controlled by the IR Green's function as that at relativistic fermionic fixed point.
Further, in \cite{Li:2011nz,Kuang:2012tq}, they study the non-relativistic fermionic spectrum in the presence of dipole coupling
and can't observe the emergence of gap up to $p=8$.

In this letter, we shall study the properties of fermionic response from BI AdS black hole,
which is the corrections to the Maxwell sector.
It is the first time to study the effects on fermionic spectrum from the corrections of the gauge field sector.
The BI action is a non-linear generalization of the Maxwell theory \cite{Born:1934}
\begin{eqnarray}
\label{LM}
\mathcal{L}_{BI}=\frac{4}{\gamma}\left(1-\sqrt{1+\frac{\gamma}{2}F^2}\right).
\end{eqnarray}
The replace of Maxwell action by the BI action is natural in string theory \cite{Gibbons:2001gy}.
The nonlinearity of BI action is controlled by the BI parameter $\gamma$,
which has dimension of the square of length and is related to the string tension $\alpha'$ as $\gamma=(2\pi\alpha')^2$.
When $\gamma\rightarrow 0$, the Born-Infeld term reduces to Maxwell term, i.e., $\mathcal{L}_M=-F^2$,
whereas in the limit $\gamma\rightarrow \infty$, it vanishes.

Our letter is organized as follows.
In section \ref{BI} we present a brief review on the BI-AdS geometry and analyse its IR geometry.
And then we derive the Dirac equation in BI-AdS background and give the expressions of
relativistic and non-relativistic spectral function in section \ref{DiE}.
In section \ref{FSp}, the fermionic spectrum from BI-AdS background are numerically worked out and discussed.
Conclusions and discussion are summarized in section \ref{ConD}.

\section{Einstein-Born-Infeld black hole}\label{BI}

The BI-AdS geometry and the extended studies have been explored
in detail in \cite{Dey:2004yt,Cai:2004eh,Cai:2008in,Banerjee:2011cz,Liu:2011cu,Chaturvedi:2015hra} and references therein.
Here, we only give a brief review on the BI-AdS geometry related with our present study.

We first start with the action
\begin{eqnarray}
\label{action}
S=\frac{1}{2\kappa^2}\int d^{4}x \sqrt{-g}
\left[R+\frac{6}{L^2}
+\mathcal{L}_{BI}
\right]\,.
\end{eqnarray}
This action supports a charged BI-$AdS_4$ black hole solution \cite{Dey:2004yt,Cai:2004eh},
\begin{eqnarray}
\label{MetricR}
&&
ds^{2}=-g_{tt}dt^{2}+g_{uu}du^2+g_{xx}dx^2+g_{yy}dy^2,
\\
&&
\label{gR}
g_{tt}=\frac{f(u)}{u^2},~~~~g_{uu}=\frac{1}{u^2f(u)},~~~~g_{xx}=g_{yy}=\frac{1}{u^2},
\\
&&
\label{frR}
f(u)=1
-Mu^3
+\frac{2}{3\gamma}\left(1-\sqrt{1+\gamma Q^2 u^4}\right)
+\frac{4Q^2u^4}{3}{}_2F_{1}[\frac{1}{2},\frac{1}{4};\frac{5}{4};-\gamma Q^2u^4],
\\
&&
\label{AaR}
A_t=\mu\left({}_2F_{1}[\frac{1}{2},\frac{1}{4};\frac{5}{4};-\gamma\mu^2]
-u~{}_2F_{1}[\frac{1}{2},\frac{1}{4};\frac{5}{4};-\gamma\mu^2 u^4]\right),
\\
&&
\label{muQR}
M=1
+\frac{2}{3\gamma}\left(1-\sqrt{1+\gamma Q^2}\right)
+\frac{4Q^2}{3}{}_2F_{1}[\frac{1}{2},\frac{1}{4};\frac{5}{4};-\gamma Q^2],~~~~\mu=Q,
\end{eqnarray}
where ${}_2F_{1}[a,b;c;z]$ is a hypergeometric function.
The horizon locates at $u=1$ and the boundary at $u=0$. The dimensionless temperature is given by
\begin{eqnarray}
\label{tem}
T=\frac{1}{4\pi}\Big[3+\frac{2}{\gamma}(1-\sqrt{1+\gamma Q^2})\Big]
.
\end{eqnarray}
Note that in the limit $\gamma\rightarrow 0$, the redshift factor $f(u)$ and the gauge field $A_t(u)$ reduce to that of RN-AdS, respectively.

Before proceeding, we shall analyse the IR geometry of BI black hole,
which is important to understand the low frequency behavior of holographic fermionic spectrum.
Here we only focus on the zero temperature limit, which is obtained by setting
\begin{eqnarray}
\label{mugamma}
\mu=\mu_{\gamma}\equiv\frac{\sqrt{9\gamma+12}}{2}\,.
\end{eqnarray}
In this case, the redshift factor $f(u)$ becomes
\begin{eqnarray}
\label{frtem0r1}
f(u)|_{T=0,u\rightarrow 1}\simeq \frac{3(4+3\gamma)}{2+3\gamma}(u-1)^2
\equiv\frac{1}{L_2^2}(u-1)^2,
\end{eqnarray}
where we have defined $L_2\simeq\sqrt{\frac{2+3\gamma}{3(4+3\gamma)}}$,
which is explicitly dependent on the BI parameter $\gamma$.
Considering the following scaling limit
\begin{eqnarray} \label{ScalingLimitAdS2}
u-1=-\epsilon \frac{L_{2}^{2}}{\varsigma}~,~~~
t=\epsilon^{-1}\tau~,~~~~\epsilon\rightarrow 0,~~with~~\varsigma, \tau~~finite,
\end{eqnarray}
under which, the near horizon metric and gauge field can be wrote as
\begin{eqnarray} \label{MetricNearHorizon}
ds^{2}=\frac{L_{2}^{2}}{\varsigma^{2}}(-d\tau^{2}+d\varsigma^{2})+dx^{2}+dy^2~,
~~~~
A_{\tau}=\frac{e}{\varsigma}~,
\end{eqnarray}
with $e=\frac{\mu}{\sqrt{1+\gamma\mu^2}} L_2^2$.
Therefore, as that of RN-AdS geometry, the near horizon geometry of BI AdS black hole is $AdS_2\times \mathbb{R}^2$ with curvature radius $L_2$.

\section{Dirac equation}\label{DiE}

Subsequently we shall use the following fermion action to probe the BI-AdS geometry
\begin{eqnarray}
\label{actionspinor} S_{D}=i\int d^{4}x
\sqrt{-g}\,\overline{\zeta}\left(\Gamma^{a}\mathcal{D}_{a} - m -i
p \slashed{F}\right)\zeta,
\end{eqnarray}
where $\mathcal{D}_{a}=\partial_{a}+\frac{1}{4}(\omega_{\mu\nu})_{a}\Gamma^{\mu\nu}-iq A_{a}$ and
$\slashed{F}=\frac{1}{2}\Gamma^{\mu\nu}(e_\mu)^a(e_\nu)^bF_{ab}$
with $(e_{\mu})^{a}$ and $(\omega_{\mu\nu})_{a}$ being a set of
orthogonal normal vector bases and the spin connection 1-forms, respectively.
$p$ is the dipole coupling strength.

Making a redefinition of spinor field $\zeta=(g_{tt}g_{xx}g_{yy})^{-\frac{1}{4}}\mathcal{F}$
and the Fourier expansion with $k_x=k$ and $k_y=0$,
\begin{eqnarray}
\mathcal{F}=\int\frac{d\omega dk}{2\pi}F(u,k)e^{-i\omega t + ikx},
\end{eqnarray}
the Dirac equation can be deduced from the above action (\ref{actionspinor}) as following
\begin{eqnarray} \label{DiracEF}
\left[(\partial_{u}-m\sqrt{g_{uu}}\sigma^3)
+\sqrt{\frac{g_{uu}}{g_{tt}}}(\omega+qA_{t})i\sigma^2
+((-1)^{I} k \sqrt{\frac{g_{uu}}{g_{xx}}}
+p\sqrt{g^{tt}}\partial_{u}A_{t})\sigma^1 \right] F_{I} =0~,
\end{eqnarray}
with $I=1,2$.
In the above equations, we have used the following gamma matrices
\begin{eqnarray}
 && \Gamma^{u} = \left( \begin{array}{cc}
-\sigma^3 & 0  \\
0 & -\sigma^3
\end{array} \right), \;\;
 \Gamma^{t} = \left( \begin{array}{cc}
 i \sigma^1 & 0  \\
0 & i \sigma^1
\end{array} \right),  \;\;
\nonumber
\\
&&
\label{GammaMatrices}
\Gamma^{x} = \left( \begin{array}{cc}
-\sigma^2 & 0  \\
0 & \sigma^2
\end{array} \right),\;\;
\Gamma^{y} = \left( \begin{array}{cc}
0 & -i\sigma^2  \\
i\sigma^2 & 0
\end{array} \right).\;\;
\end{eqnarray}
Furthermore, we can also express the above Dirac equations in terms of 4-component spinors $\mathcal{A}_{I}$ and $\mathcal{B}_{I}$ defined as
$
F_{I} \equiv (\mathcal{A}_{I}, \mathcal{B}_{I})^{T}
$
,
\begin{eqnarray} \label{DiracEAB1}
&& (\partial_{u}-m\sqrt{g_{uu}})\mathcal{A}_{I}
+\sqrt{\frac{g_{uu}}{g_{tt}}}(\omega+qA_{t})\mathcal{B}_{I}
+((-1)^{I} k \sqrt{\frac{g_{uu}}{g_{xx}}}
+p\sqrt{g^{tt}}\partial_{u}A_{t})\mathcal{B}_{I} =0~,
\\
&& \label{DiracEAB2} (\partial_{u}+m\sqrt{g_{uu}})\mathcal{B}_{I}
-\sqrt{\frac{g_{uu}}{g_{tt}}}(\omega+qA_{t})\mathcal{A}_{I}
+((-1)^{I} k \sqrt{\frac{g_{uu}}{g_{xx}}}
+p\sqrt{g^{tt}}\partial_{u}A_{t}) \mathcal{A}_{I} =0~.
\end{eqnarray}
It is more convenient to implement the numerical computation by packaging
the above Dirac equations into the following flow equation
\begin{eqnarray} \label{DiracEF1}
(\partial_{u}-2m\sqrt{g_{uu}}) \xi_{I}
+\left[ v_{-} + (-1)^{I} k \sqrt{\frac{g_{uu}}{g_{xx}}}  \right]
+ \left[ v_{+} - (-1)^{I} k \sqrt{\frac{g_{uu}}{g_{xx}}}  \right]\xi_{I}^{2}
=0
~,
\end{eqnarray}
where we have defined $\xi_{I}\equiv \frac{\mathcal{A}_{I}}{\mathcal{B}_{I}}$
and $v_{\pm}=\sqrt{\frac{g_{uu}}{g_{tt}}}(\omega+q
A_{t})\mp p \sqrt{g^{tt}}\partial_{u}A_{t}$.
To solve the above flow equation, we shall impose the boundary conditions at the horizon $u=1$ for $\omega\neq
0$,
\begin{eqnarray} \label{BCu0}
\xi_I|_{u=1,\omega\neq0}=i\,,
\end{eqnarray}
which is based on the requirement of ingoing wave propagating near the horizon.
While for $T=0$ and $\omega=0$, an alternative boundary condition at $u=1$ should be imposed as \cite{Liu:2009dm,Wu:2011bx}
\begin{eqnarray} \label{BCu1}
\xi_I|_{u=1,\omega=0}=\frac{m-\sqrt{k^2+m^2-(q\mu_{\gamma}L_2)^2-i\epsilon}}{k+q\mu_{\gamma}L_2}\,.
\end{eqnarray}

Once we have the flow equation (\ref{DiracEF1}) with the boundary conditions at the horizon in hand,
we can read off the boundary Green's function following the prescriptions in \cite{Faulkner:2009wj},
\begin{eqnarray} \label{Green}
G (\omega,k)= \lim_{u\rightarrow \infty} u^{-2m}
\left( \begin{array}{cc}
\xi_{1}   & 0  \\
0  & \xi_{2} \end{array} \right)  \ .
\end{eqnarray}
It is the case of the relativistic fixed point \cite{Liu:2009dm},
in which the bulk action (\ref{actionspinor}) is accompanied by a Lorentz covariance boundary term as
\begin{eqnarray} \label{SbdyL}
S_{bdy}=\frac{i}{2}\int_{\partial \mathcal{M}}d^3x\sqrt{-h}\bar{\zeta}\zeta\,,
\end{eqnarray}
where $h$ is the determinant of induced metric on the boundary.
And then the spectral function is defined as
\begin{eqnarray}
A(\omega,k)\equiv \text{ImTr} G(\omega,k)=\text{Im}[G_{11}(\omega,k)+G_{22}(\omega,k)]\,.
\end{eqnarray}
But from the Dirac flow equation (\ref{DiracEF1}) it is easy to infer that
\begin{eqnarray}\label{G22G11}
G_{22}(\omega,k)=G_{11}(\omega,-k)\,.
\end{eqnarray}
Therefore, at the relativistic fermionic fixed point, we usually focus on $G_{22}(\omega,k)$ instead of $A(\omega,k)$.

On the other hand, if we replace the boundary term (\ref{SbdyL}) by a Lorentz violating one \cite{Laia:2011wf,Laia:2011zn}
\begin{eqnarray} \label{SbdyLN}
S_{bdy}=\frac{1}{2}\int_{\partial \mathcal{M}}d^3x\sqrt{-h}\bar{\zeta}\Gamma^x\Gamma^y\zeta\,,
\end{eqnarray}
we shall have a non-relativistic fixed point in which the dual field theory is not Lorentz covariant.
In this case, the fermionic spectral function can be expressed in terms of the retarded function at the relativistic fixed point \cite{Laia:2011zn,Li:2011nz}
\begin{eqnarray} \label{GreenNR}
G_{NR}=
\left( \begin{array}{cc}
\frac{2G_{11}G_{22}}{G_{11}+G_{22}}   & \frac{G_{11}-G_{22}}{G_{11}+G_{22}}  \\
\frac{G_{11}-G_{22}}{G_{11}+G_{22}}  & \frac{-2}{G_{11}+G_{22}} \end{array} \right) \,.
\end{eqnarray}
Consequently, the spectral function at non-relativistic fixed point has the form
\begin{eqnarray} \label{ANRs}
A_{NR}(\omega,k)=
\text{ImTr} [G_{NR}]=\text{Im}\Big[\frac{2G_{11}G_{22}-2}{G_{11}+G_{22}}\Big]\,.
\end{eqnarray}
In this letter, we shall discuss the fermionic spectrum dual to BI gravity at relativistic fixed point and non-relativistic one, respectively.

\section{Relativistic fermionic spectrum}\label{FSp}

In this section, we shall systematically study the fermionic spectrum dual to BI-AdS geometry by numerically solving the Dirac equations.

\subsection{Fermionic spectrum without dipole coupling}

\begin{figure}
\center{\includegraphics[scale=0.3]{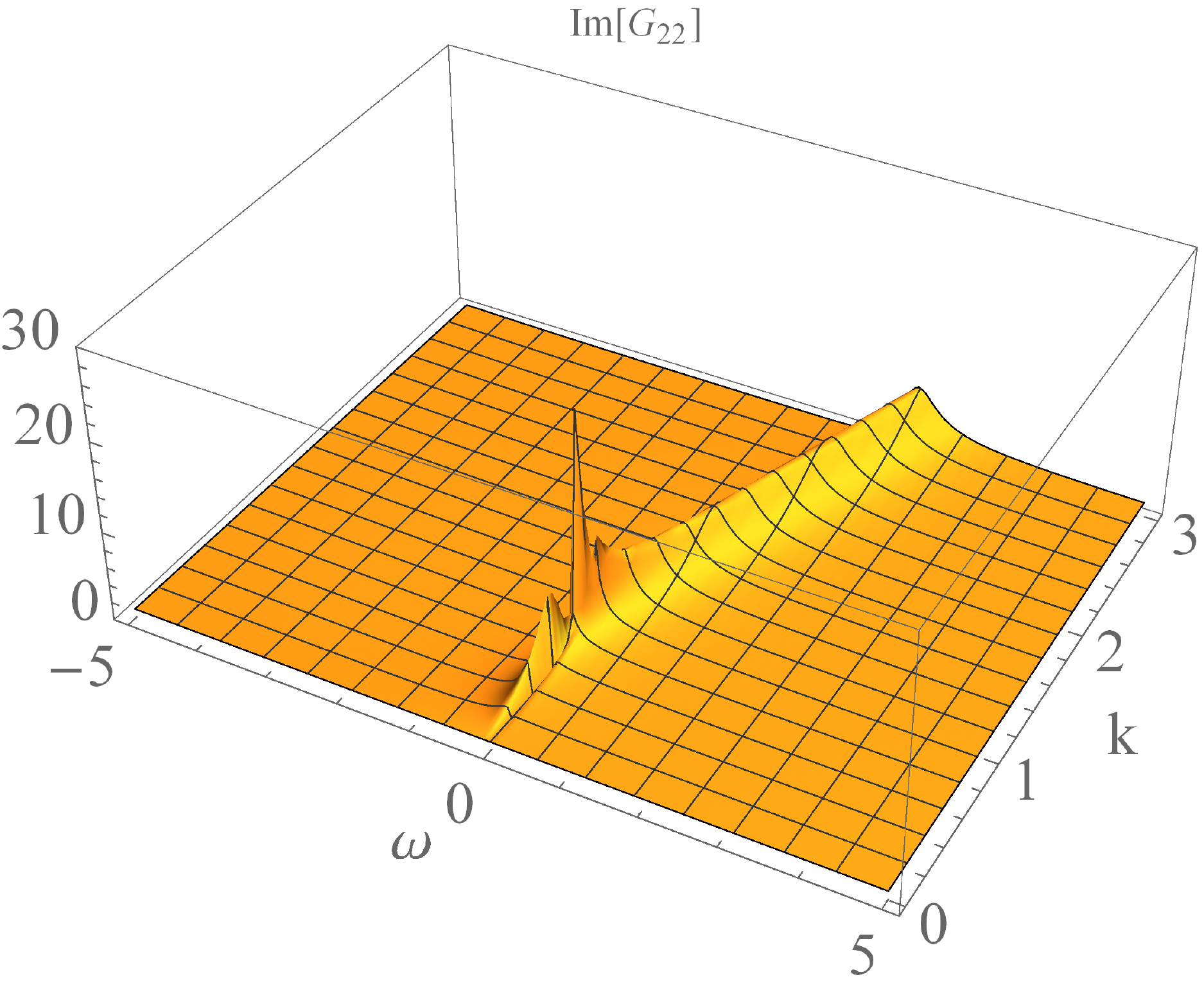}\ \hspace{0.6cm}
\includegraphics[scale=0.26]{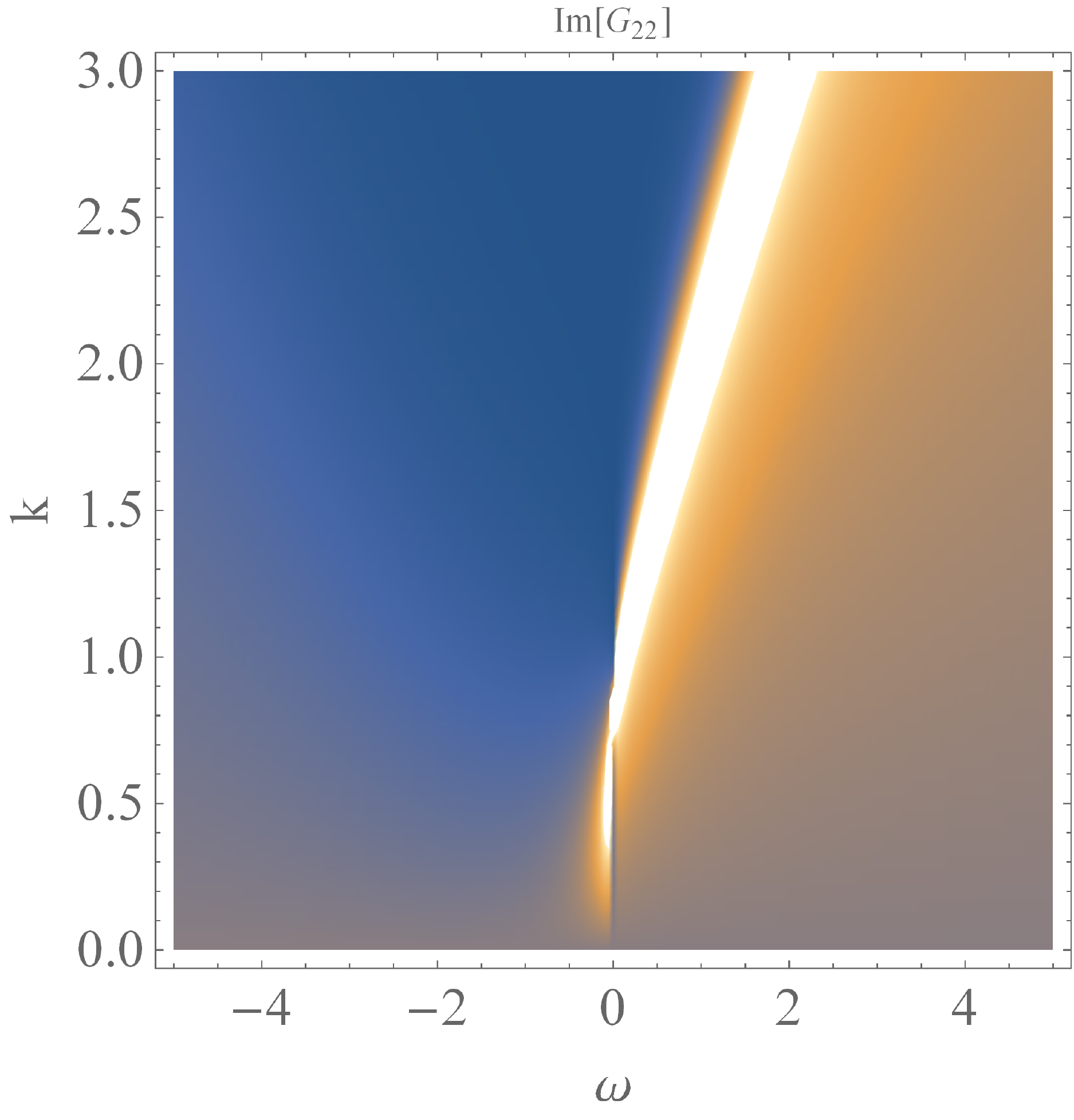}\ \\
\caption{\label{3Dgamma0p4p0}The 3d and density plots of Im$G_{22}(\omega,k)$ for $\gamma=0.4$.
A sharp quasi-particle-like peak can be observed around $\omega=0$ and $k_F\simeq 0.83$.
All the other parameters are fixed as $q=1$, $m=0$, $T=0$ and $p=0$.}}
\end{figure}

In this subsection, we explore the fermionic spectrum without dipole coupling term at relativistic fixed point.
Our interesting point mainly focus on how does the BI parameter $\gamma$ affect the fermionic spectrum
and so we fit $q=1$, $m=0$, $T=0$ and $p=0$ without any real loss of generality.
Similar with that in RN-AdS black hole \cite{Liu:2009dm}, a sharp quasi-particle-like peak near $\omega=0$ and $k_F\simeq 0.83$
can also be found in the holographic fermionic spectrum dual to BI-AdS black hole (FIG.\ref{3Dgamma0p4p0}).
Next, we shall quantitatively study the relation between the Fermi momentum $k_F$ and the BI parameter $\gamma$.

\begin{figure}
\center{\includegraphics[scale=0.54]{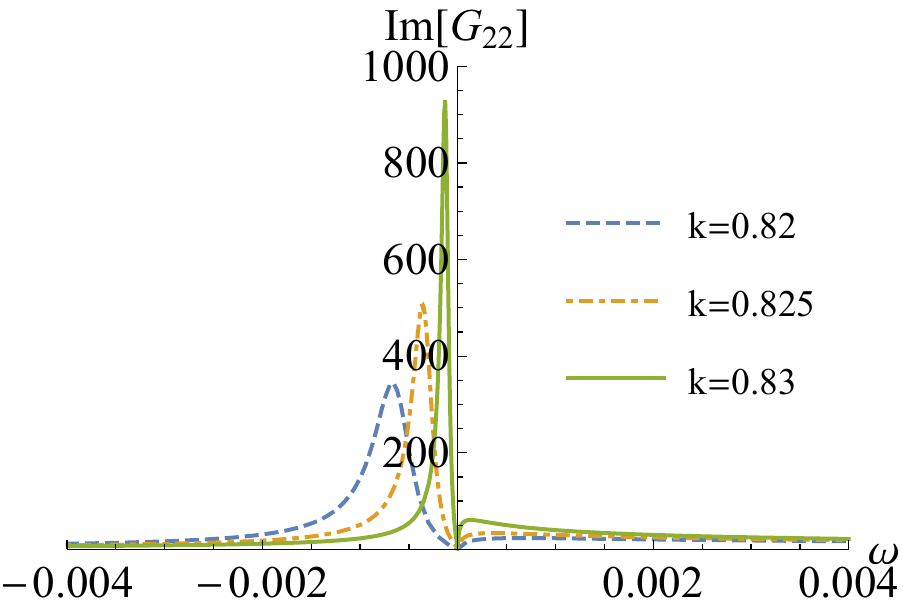}\ \hspace{0.2cm}
\includegraphics[scale=0.54]{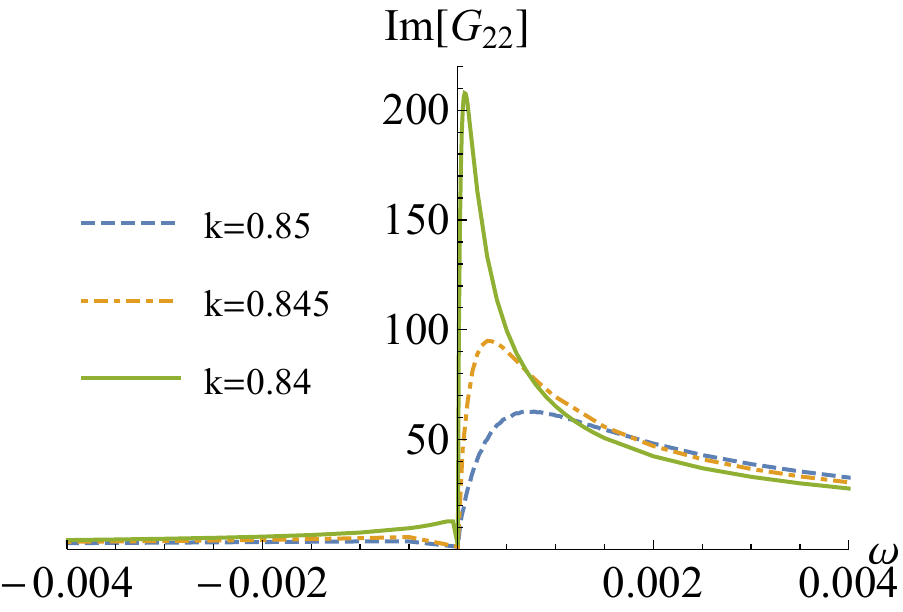}\ \hspace{0.2cm}
\includegraphics[scale=0.54]{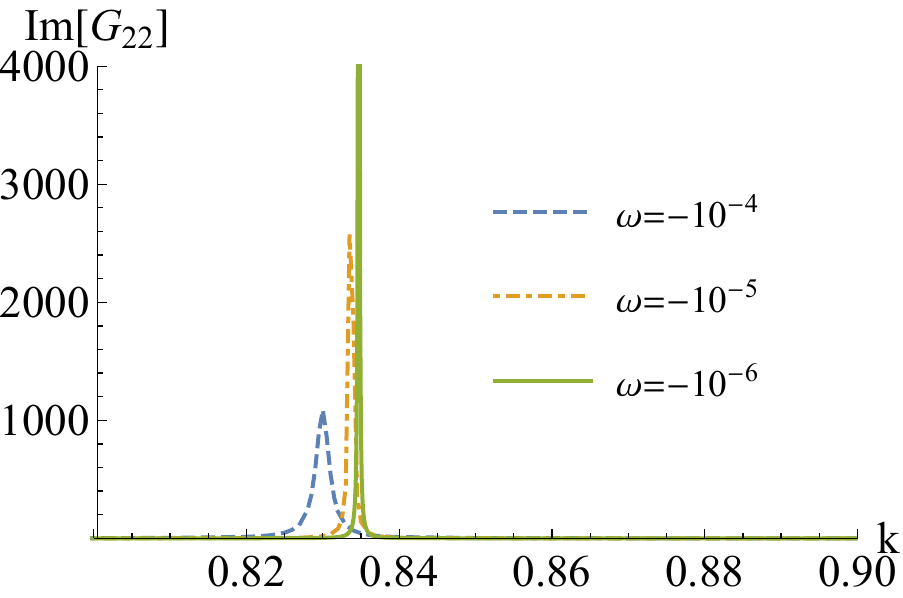}
\caption{\label{ImG22}The first and second plots: Im$G_{22}(\omega,k)$ as a function of $\omega$ for a given $k$.
The third plot: Im$G_{22}(\omega,k)$ as a function of $k$ for a given $\omega$.
All the other parameters are fixed as $\gamma=0.4$, $q=1$, $m=0$, $T=0$ and $p=0$.}}
\end{figure}

Before proceeding, we shall follow the procedure in \cite{Liu:2009dm} to demonstrate that the quasi-particle-like peak observed in FIG.\ref{3Dgamma0p4p0}
is an infinitely sharp excitations.
To this end, we focus the behavior in the region of small $k_{\perp}\equiv k-k_F$ and $\omega$.
Firstly, we show Im$G_{22}(\omega,k)$ as a function of $\omega$ for a given $k$ and $\gamma=0.4$ in the first and second plots in FIG.\ref{ImG22}.
As $k_{\perp}\rightarrow 0_-$, both peak located in the region $\omega<0$ and bump in $\omega>0$ approach $\omega=0$ (the first plot in FIG.\ref{ImG22})\footnote{When $k_{\perp}\rightarrow 0_+$,
both bumps approach $\omega=0$ (see the second plot in FIG.\ref{ImG22}).}.
Eventually, they meet and produce infinitely sharp excitations with infinite heights and zero widths near $\omega=0$ and $k=k_F$.
On the other hand, we plot Im$G_{22}(\omega,k)$ as a function of $k$ for a given small $\omega$ in the third panel in FIG.\ref{ImG22}.
We find that in the limit $\omega\rightarrow 0_-$, a sharp excitation with infinite height and zero width produces, implying that there is an Fermi peak located near $\omega=0$.
We can work out the location of Fermi peak in the momentum space in the limit $\omega\rightarrow 0_-$ to get the Fermi momentum as $k_F\simeq 0.8352$ for $\gamma=0.4$.

In determining the Fermi momentum $k_F$, there are some subtleties. We present as follows.
Firstly, for $\omega=0$ and $m=0$, if $k>q\mu_{\gamma}L_2$, the boundary condition (\ref{BCu1}) at $u=1$ is real.
Together with the real Dirac equation (\ref{DiracEF1}), we can deduce that the Im$G_{11}$ and Im$G_{22}$ are identically zero for $k>q\mu_{\gamma}L_2$ at $\omega=0$.
For instance, when $\gamma=0$, which reduces to the case of RN-AdS, $k_F\simeq 0.9185$ is belong to the region $k>\frac{q\mu}{\sqrt{6}}=\frac{1}{\sqrt{2}}$.
So in this case we cannot impose the boundary condition (\ref{BCu1}).
Alternatively, we should impose the boundary condition (\ref{BCu0}) in the limit $\omega\rightarrow 0_-$ to locate the Fermi momentum $k_F$.
Secondly, since the Dirac equation (\ref{DiracEF1}) is singular at the horizon $u=1$, in numerics the boundary condition must be impose close to the horizon instead of the horizon itself.
Also, to see the infinitely sharp excitations, we have to impose the boundary condition very close to the horizon.
Here, we impose the boundary condition at $u=1-10^{-6}$.

\begin{figure}
\center{\includegraphics[scale=0.6]{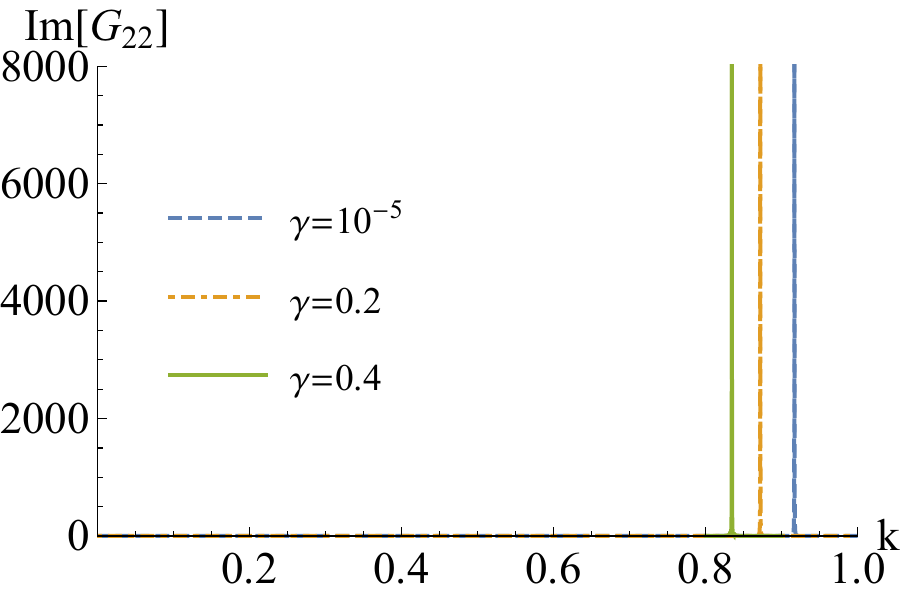}\ \hspace{0.4cm}
\includegraphics[scale=0.46]{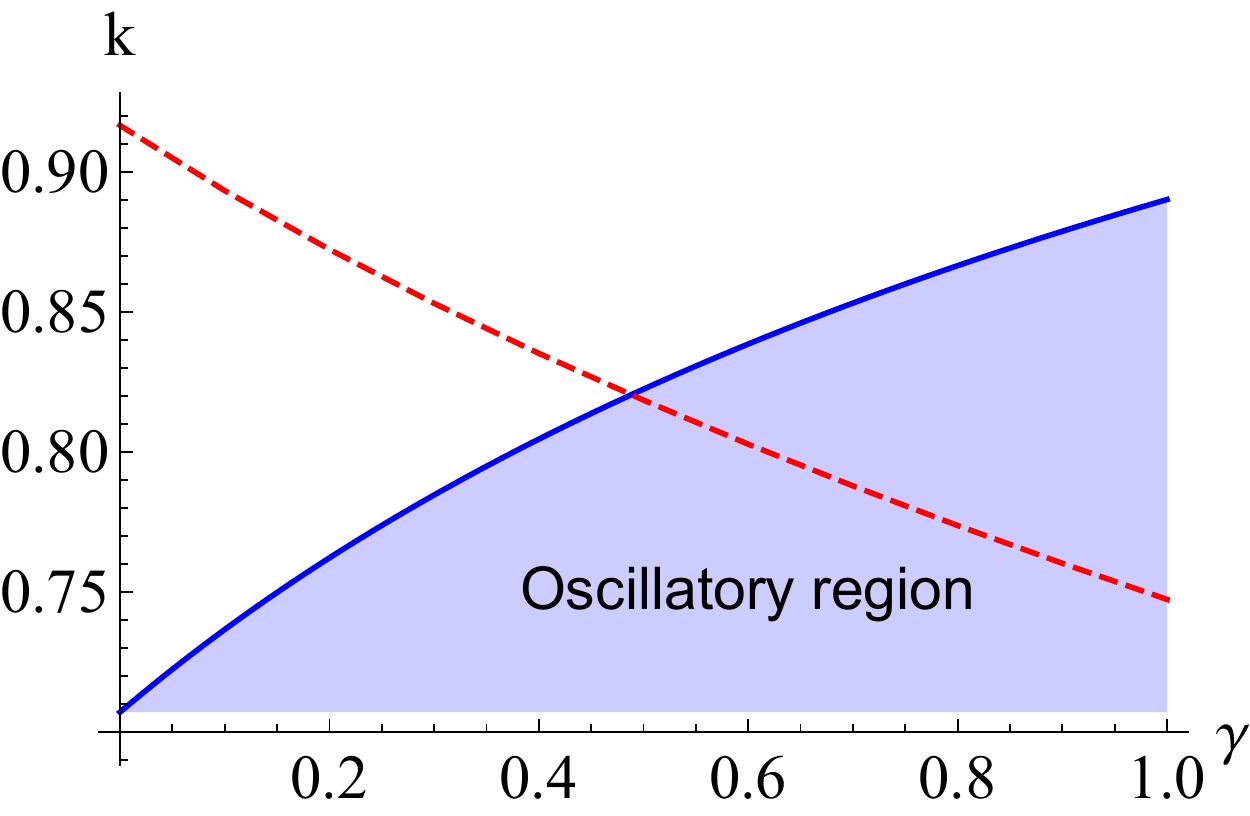}
\caption{\label{kFvsgamma}Left plot: Im$G_{22}$ as a function of $k$ at $\omega=-10^{-6}$ for sample BI parameter $\gamma$.
Right plot: The red dashed line is the relation between the BI parameter $\gamma$ and the location of the peak of Im$G_{22}$ as a function of $k$ for $\omega=-10^{-6}$.
The blue zone is the oscillatory region.
All the other parameters are fixed as $q=1$, $m=0$, $T=0$ and $p=0$.}}
\end{figure}

Based on the above prescription on the excitation of Fermi peak and the key points on the numerics,
we plot the Im$G_{22}$ as a function of $k$ at $\omega=-10^{-6}$ for sample BI parameter $\gamma$ in the left plot in FIG.\ref{kFvsgamma}.
We can see that with increase of $\gamma$, Fermi momentum $k_F$ decreases.
Further, we show the relation between $\gamma$ and the location of the peak of Im$G_{22}$ as a function of $k$ for $\omega=-10^{-6}$ in the right plot in FIG.\ref{kFvsgamma}.
The blue zone is the oscillatory region, in which $\nu_I(k)$ becomes pure imaginary\footnote{$\nu_I(k)$ is defined as $\nu_I(k)\equiv\sqrt{(m^2+k^2)L_2^2-q^2e^2}$,
which relates the conformal dimension of the dual operator in the IR CFT as $\delta_k=\frac{1}{2}+\nu_I(k)$. When $\nu_I(k)$ becomes pure imaginary, the UV Green's function is periodic
in log$\omega$ and so the region of $k$ satisfying $\nu_I(k)$ being pure imaginary is dubbed as the oscillatory region. For more details, please see \cite{Liu:2009dm,Faulkner:2009wj,Wu:2013xta}.}.
Outside the oscillatory region, the peak signals a Fermi surface (right plot in FIG.\ref{ImG22}).
While when the peak enter the oscillatory region, it loses its meaning as Fermi surface.

Once the Fermi momentum $k_F$ is worked out, we can analytically obtain the scaling exponent $z$ of dispersion relation\footnote{We have used the analytical expression of
the scaling exponent $z$ (Eq.(93) in \cite{Faulkner:2009wj}), which is applicable for that with $AdS_2$ near horizon geometry.}.
Our results are summarized in Table \ref{Tfit}. We find that $z>1$ and as $\gamma$ increases, $z$ also increase.
It indicates that the holographic fermionic system dual to BI-AdS black hole is non-Fermi liquid.
Moreover, with the increase of BI parameter $\gamma$, the degree of deviation from Fermi liquid become more obvious.

\begin{widetext}
\begin{table}[ht]
\begin{center}
\begin{tabular}{|c|c|c|c|c|c|c|c|}
         \hline
~$\gamma$~ &~$10^{-5}$~&~$0.1$~&~$0.2$~&~$0.3$~&~$0.4$~
          \\
        \hline
~$k_F$~ &~$0.9124$~&~$0.8933$~&~$0.8723$~&~$0.8531$~&~$0.8352$~
          \\
        \hline
~$z$~ & ~$2.0999$~&~$2.3431$~&~$2.7151$~ & ~$3.3632$~&~$4.9262$~
          \\
        \hline
\end{tabular}
\caption{\label{Tfit}The scaling exponent $z$ with different BI parameter $\gamma$. All the other parameters are fixed as $q=1$, $m=0$, $T=0$ and $p=0$.}
\end{center}
\end{table}
\end{widetext}

\subsection{Fermionic spectrum with dipole coupling}

In this subsection, we shall turn on the dipole coupling to see
the common effects of dipole coupling $p$ and BI parameter $\gamma$
on the formation of gap.

\subsubsection{zero temperature}
\begin{figure}
\center{
\includegraphics[scale=0.3]{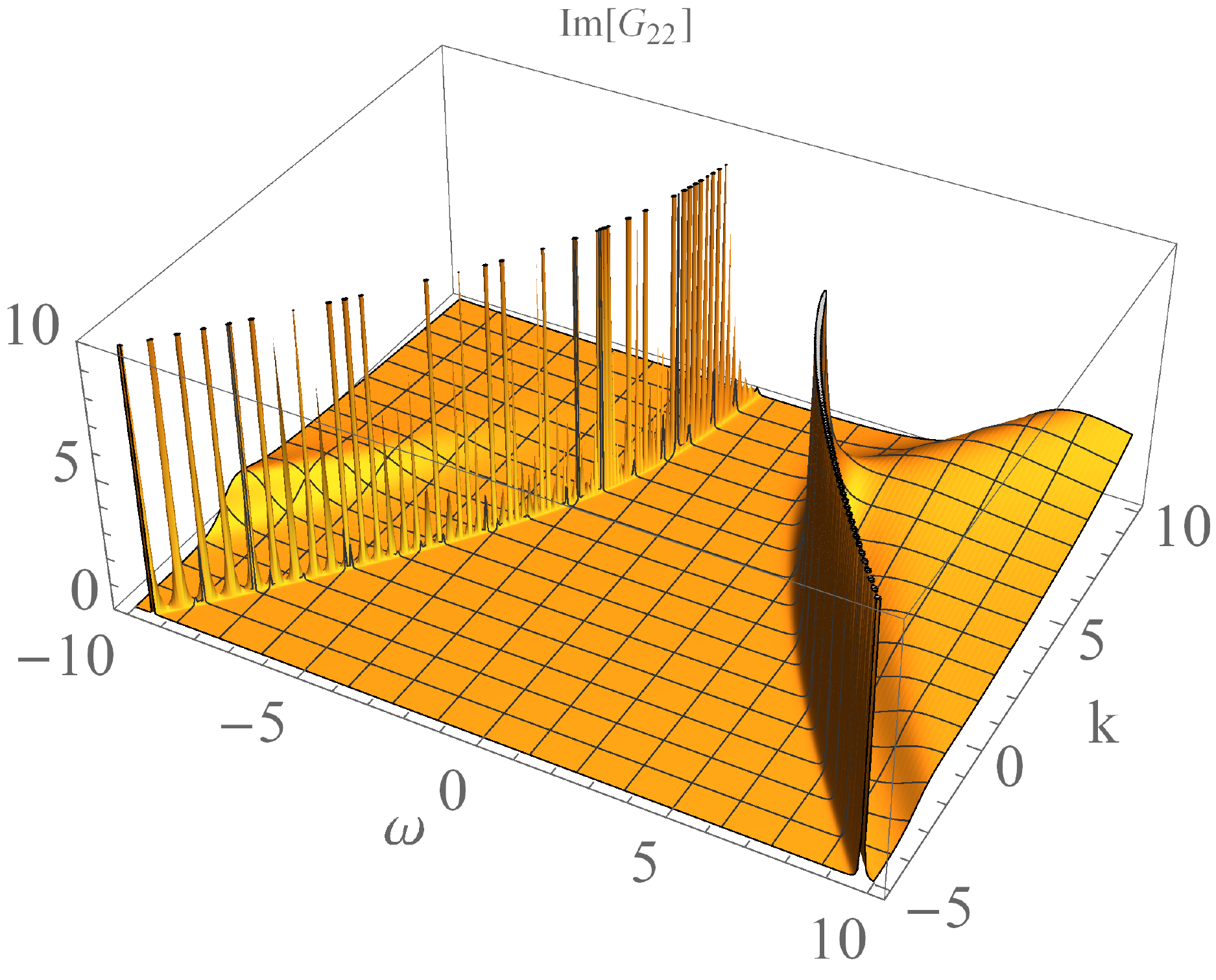}\ \hspace{0.6cm}
\includegraphics[scale=0.26]{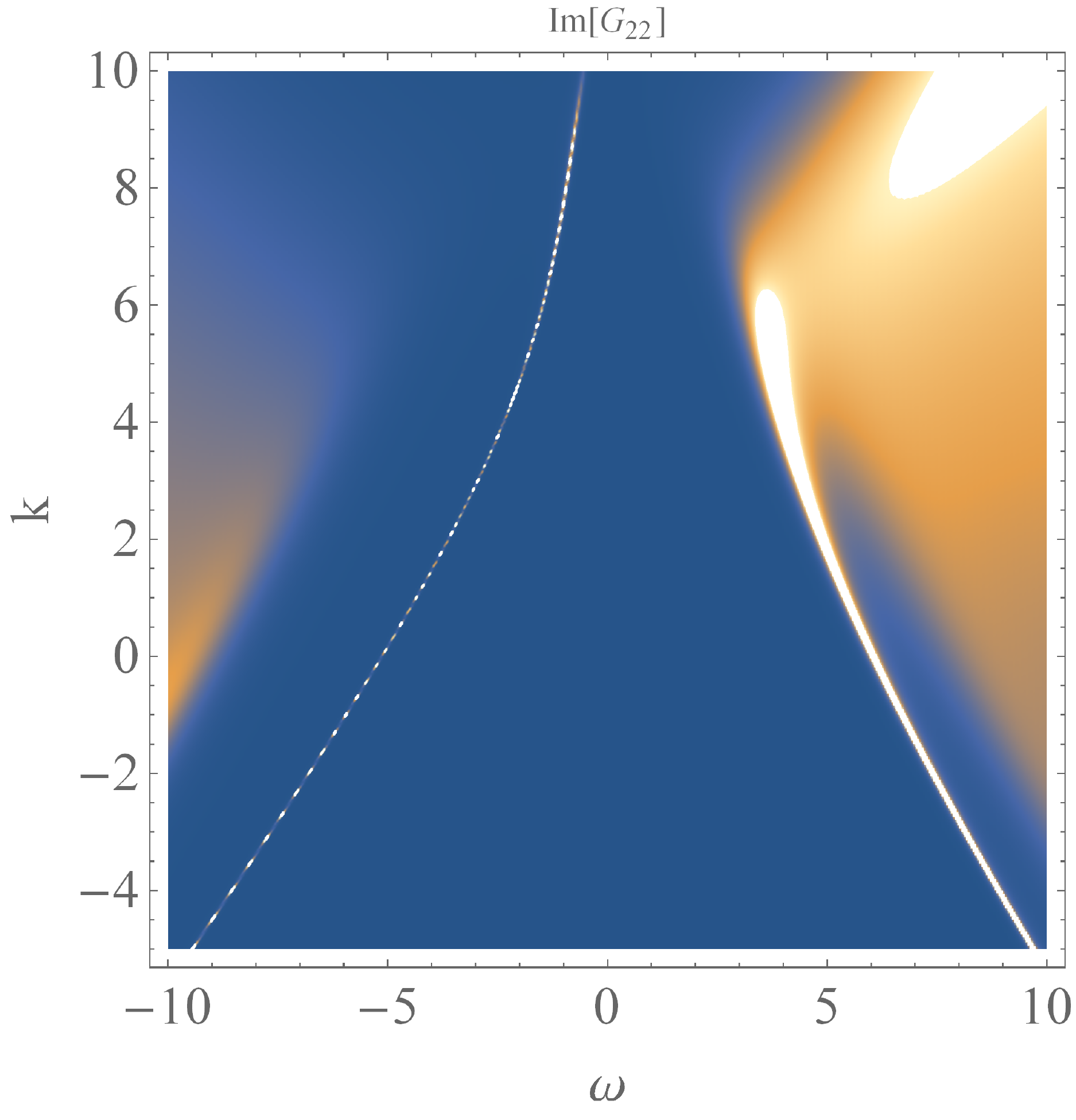}
\caption{\label{3Dgamma0p45}The 3d and density plots of Im$G_{22}(\omega,k)$ with $\gamma=0.45$ for $p=10$, in which a gape can be observed.
All the other parameters are fixed as $q=1$, $m=0$, $T=0$.}}
\end{figure}

In the last subsection, we have seen that although with the increase of BI parameter $\gamma$,
the peak of spectral function enters into the oscillatory region and loses the meaning of Fermi surface,
but Mott gap doesn't occur for $p=0$.
Therefore, we will introduce dipole coupling term between the spinor field and gauge field as in \cite{Edalati:2010ww,Edalati:2010ge}
to see the formation of Mott gap in BI-AdS background.
In particular, we will pay close attention to the effects of BI parameter $\gamma$ on the formation of Mott gap.

\begin{figure}
\center{
\includegraphics[scale=0.45]{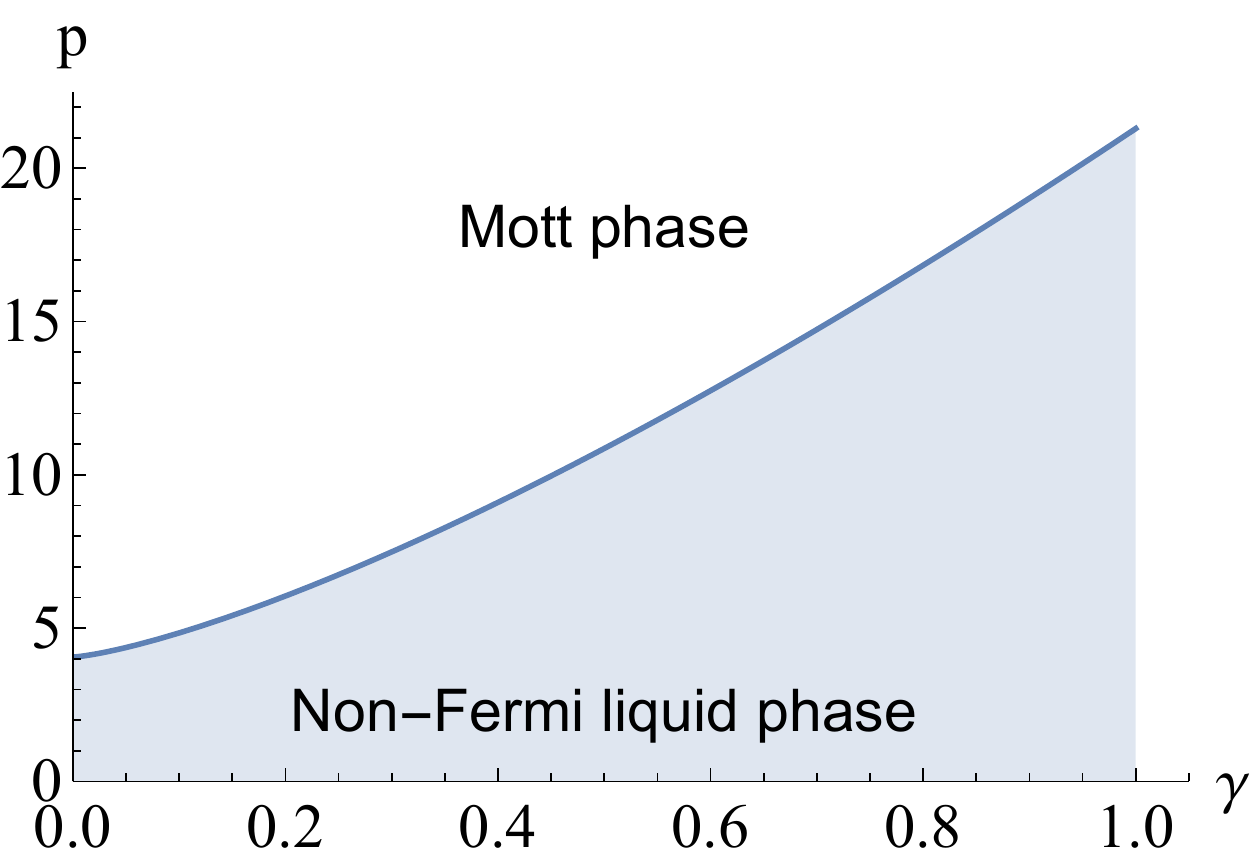}
\caption{\label{pcvsgamma}The critical value $p_c$ of Mott transition vs. $\gamma$ for
$q=1$, $m=0$ and $T=0$.}}
\end{figure}

We firstly show 3d and density plots of Im$G_{22}(\omega,k)$ with $\gamma=0.45$ for $p=10$ in FIG.\ref{3Dgamma0p45}.
A hard gap indeed emerges in the fermionic spectrum dual to BI-AdS background when dipole coupling $p$ exceeds some critical value,
which is similar with that found in RN-AdS background \cite{Edalati:2010ww,Edalati:2010ge} and other geometries \cite{Wu:2012fk,Kuang:2012ud,Kuang:2012tq,Wu:2013oea,Wu:2014rqa,Ling:2014bda,Kuang:2014yya,Vanacore:2015,Fan:2013zqa,Wen:2012ur}.
Furthermore, we show the phase diagram $(\gamma,p)$ in FIG.\ref{pcvsgamma}.
The blue line is the critical line, above which Mott gap opens\footnote{In numerics, we determine the critical line by identifying the onset of gap with
that the density of state (DOS) $A(\omega)$ drops below some small number (here, we take $10^{-2}$) at the Fermi level.}.
From this figure, we can see that for fixed $\gamma$ a phase transition happens from non-Fermi liquid phase to Mott gapped phase with the increase of $p$.
Quantitatively, with the increase of $\gamma$, the critical value of $p$ increases (FIG.\ref{pcvsgamma} and TABLE \ref{pc}). It indicates that
the BI parameter $\gamma$ plays the role of hindering the formation of Mott gap.

\begin{widetext}
\begin{table}[ht]
\begin{center}
\begin{tabular}{|c|c|c|c|c|c|c|c|c|c|}
         \hline
~$\gamma$~ &~$10^{-5}$~&~$0.1$~&~$0.2$~&~$0.4$~&~$0.6$~&~$0.8$~&~$1$~
          \\
        \hline
~$p_c$~ & ~$4$~&~$4.94$~&~$6.06$~ & ~$8.97$~&~$12.93$~&~$16.84$~&~$21.20$~
          \\
        \hline
\end{tabular}
\caption{\label{pc}The critical value $p_c$ with different BI parameter $\gamma$. All the other parameters are fixed as $q=1$, $m=0$ and $T=0$.}
\end{center}
\end{table}
\end{widetext}

\subsubsection{Finite temperature}

\begin{figure}
\center{\includegraphics[scale=0.3]{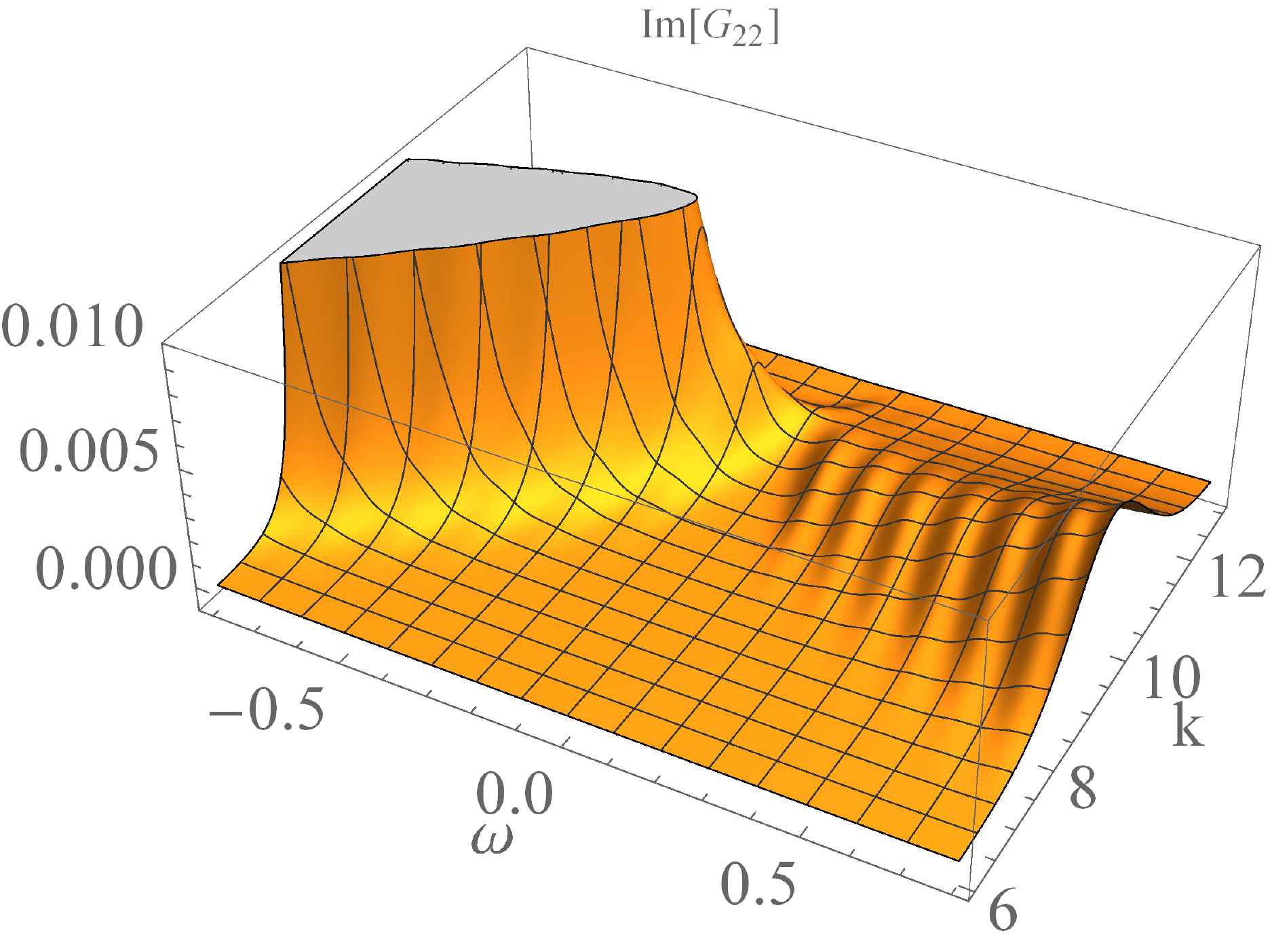}\ \hspace{0.6cm}
\includegraphics[scale=0.26]{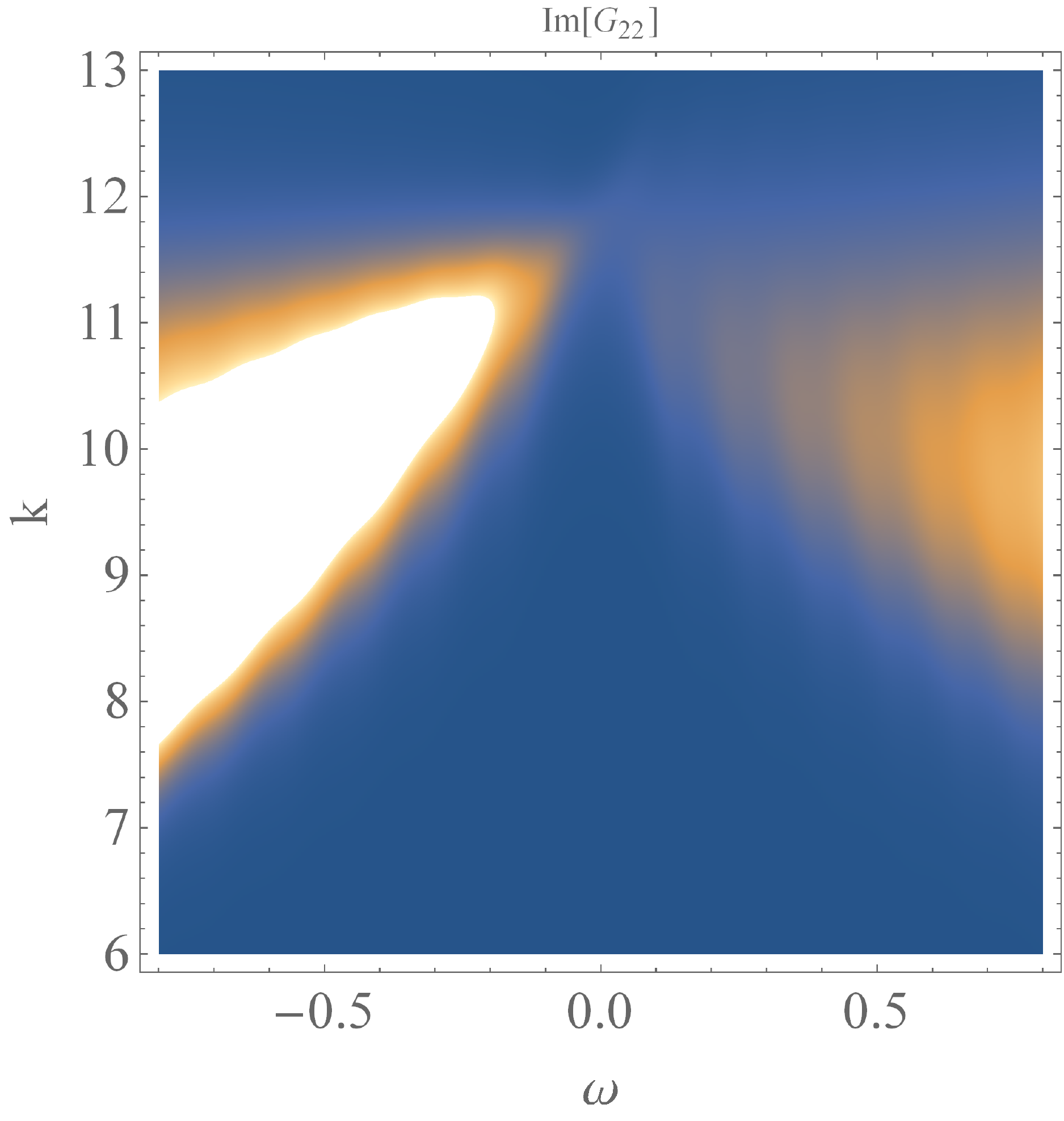}\ \\
\caption{\label{3Dgamma0p45p10T}The 3d and density plots of $ImG_{22}(\omega,k)$ for $\gamma=0.45$ and $p=10$ at $T=0.01$.
The gap closes.
All the other parameters are fixed as $q=1$ and $m=0$.}}
\end{figure}

For some Mott insulators \cite{Zylbersztejn:1975,Giamarchi:1997},
a transition from insulating phase to metallic phase happens as the temperature is increased.
The dynamics at different temperature has also been revealed in holography \cite{Edalati:2010ww,Edalati:2010ge,Ling:2014bda}.
Also they quantitatively give the ration $\Delta/T_{\ast}\simeq 10$ for $p=6$ (or $p\simeq 7$),
where $\Delta$ is the gap width at the zero temperature and $T_{\ast}$ the critical temperature at which the gap closes.
The ration $\Delta/T_{\ast}$ by holography is at the same order of magnitude as that of some transition-metal oxides such as $VO_2$,
for which it is approximately $20$. Here, we shall mainly focus on the effects of BI parameter $\gamma$ on
the dynamics at different temperature.

FIG.\ref{3Dgamma0p45p10T} shows 3d and density plots of Im$G_{22}(\omega,k)$ for $\gamma=0.45$ and $p=10$ at $T=0.01$,
in which we obviously observe that the gap closes when we heat up the system up to certain critical temperature.
Quantitatively, we present the ratio $\Delta/T_{\ast}$ for different BI parameter $\gamma$ for fixed $p=10$ in Table \ref{ratio},
from which, we can observe that with the increase of $\gamma$,
the ratio $\Delta/T_{\ast}$ decreases\footnote{Note that besides $\gamma$, the ratio $\Delta/T_{\ast}$ also depends on the other parameters in the system as $p$ and $q$.}.
\begin{widetext}
\begin{table}[ht]
\begin{center}
\begin{tabular}{|c|c|c|c|c|c|c|}
         \hline
~$\gamma$~ &~$10^{-5}$~&~$0.1$~&~$0.3$~&~$0.45$~
          \\
        \hline
~$\Delta/T_{\ast}$~ & ~$13.26$~&~$11.58$~&~$7.59$~ & ~$6.89$~
          \\
        \hline
\end{tabular}
\caption{\label{ratio}The ratio $\Delta/T_{\ast}$ with different BI parameter $\gamma$. All the other parameters are fixed as $q=1$, $m=0$ and $p=10$.}
\end{center}
\end{table}
\end{widetext}

\section{Non-relativistic fermionic spectrum}

\begin{figure}
\center{\includegraphics[scale=0.3]{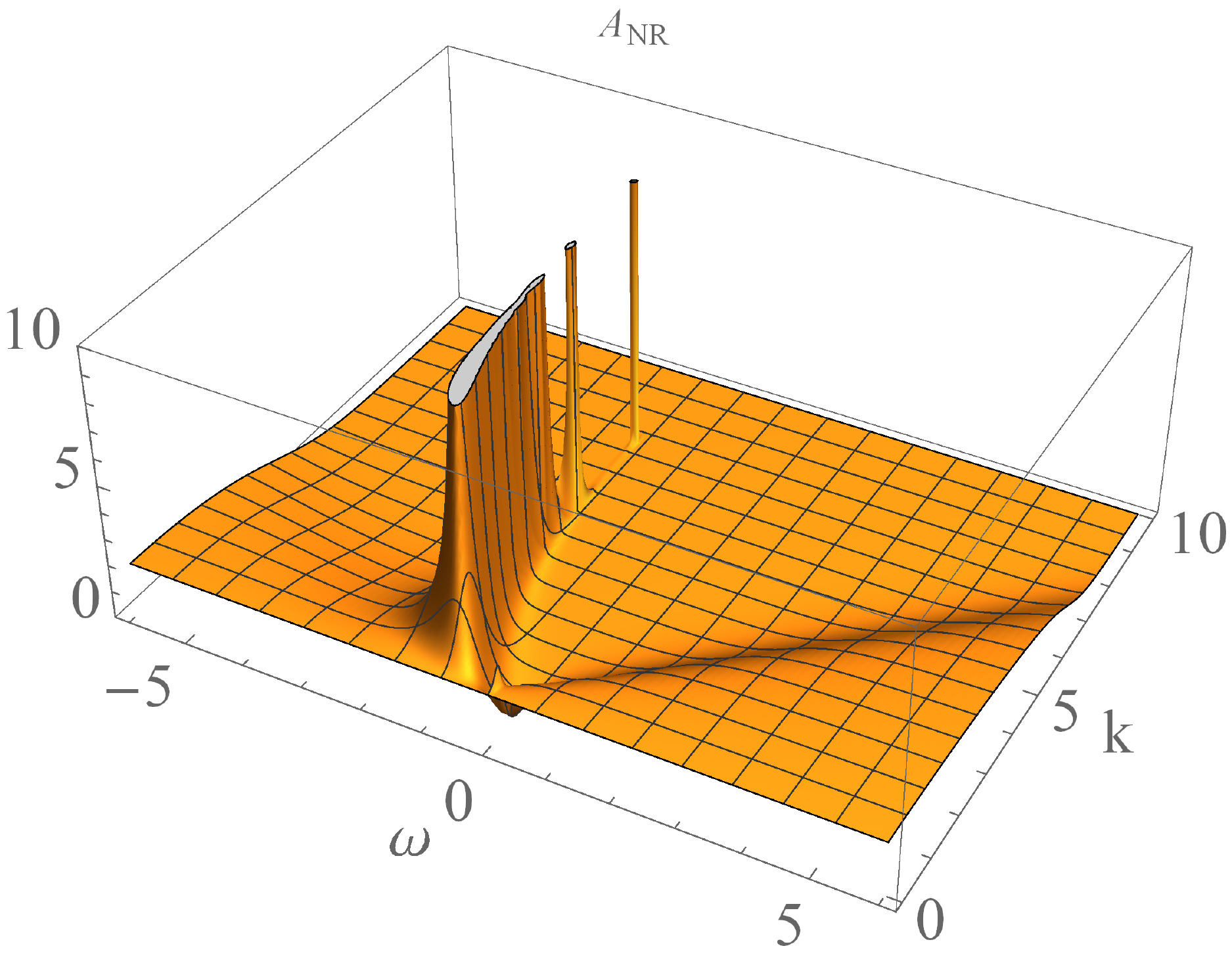}\ \hspace{0.6cm}
\includegraphics[scale=0.26]{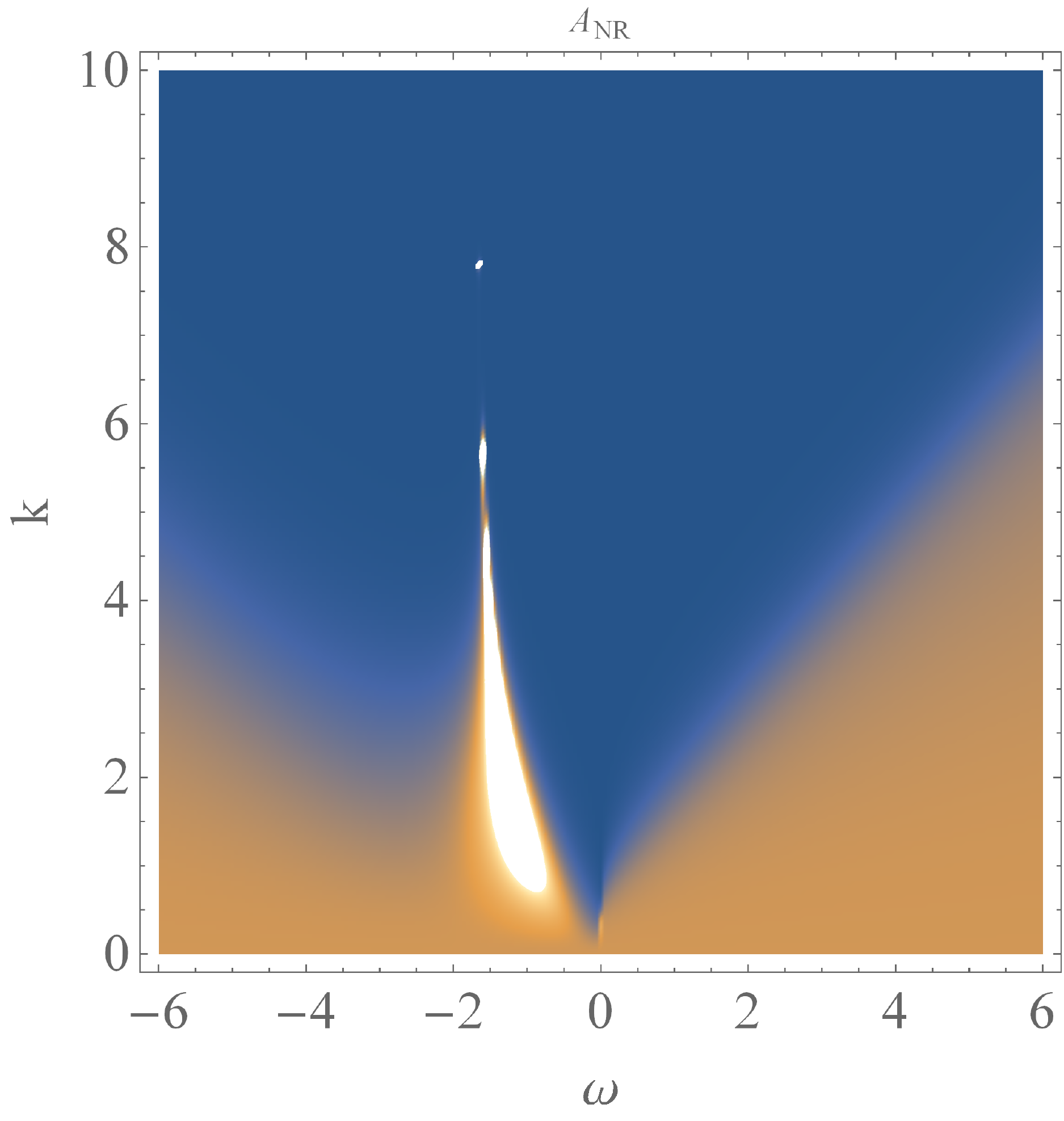}\ \\
\caption{\label{3DNR}The 3d and density plots of non-relativistic spectral function $A(\omega,k)$ for $\gamma=0.45$.
All the other parameters are fixed as $q=1$, $m=0$, $p=0$ and $T=0$.}}
\end{figure}
\begin{figure}
\center{\includegraphics[scale=0.3]{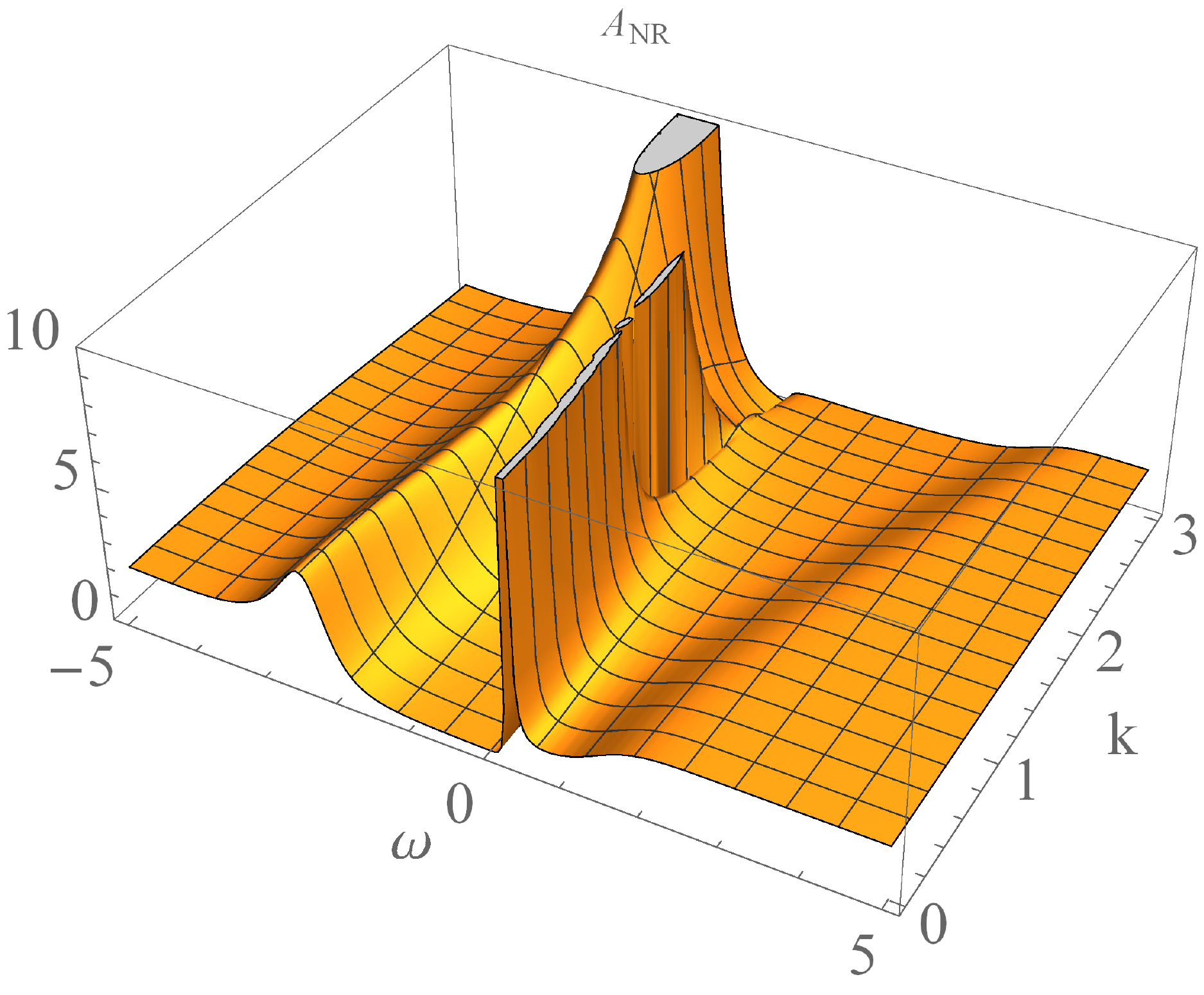}\ \hspace{0.4cm}
\includegraphics[scale=0.26]{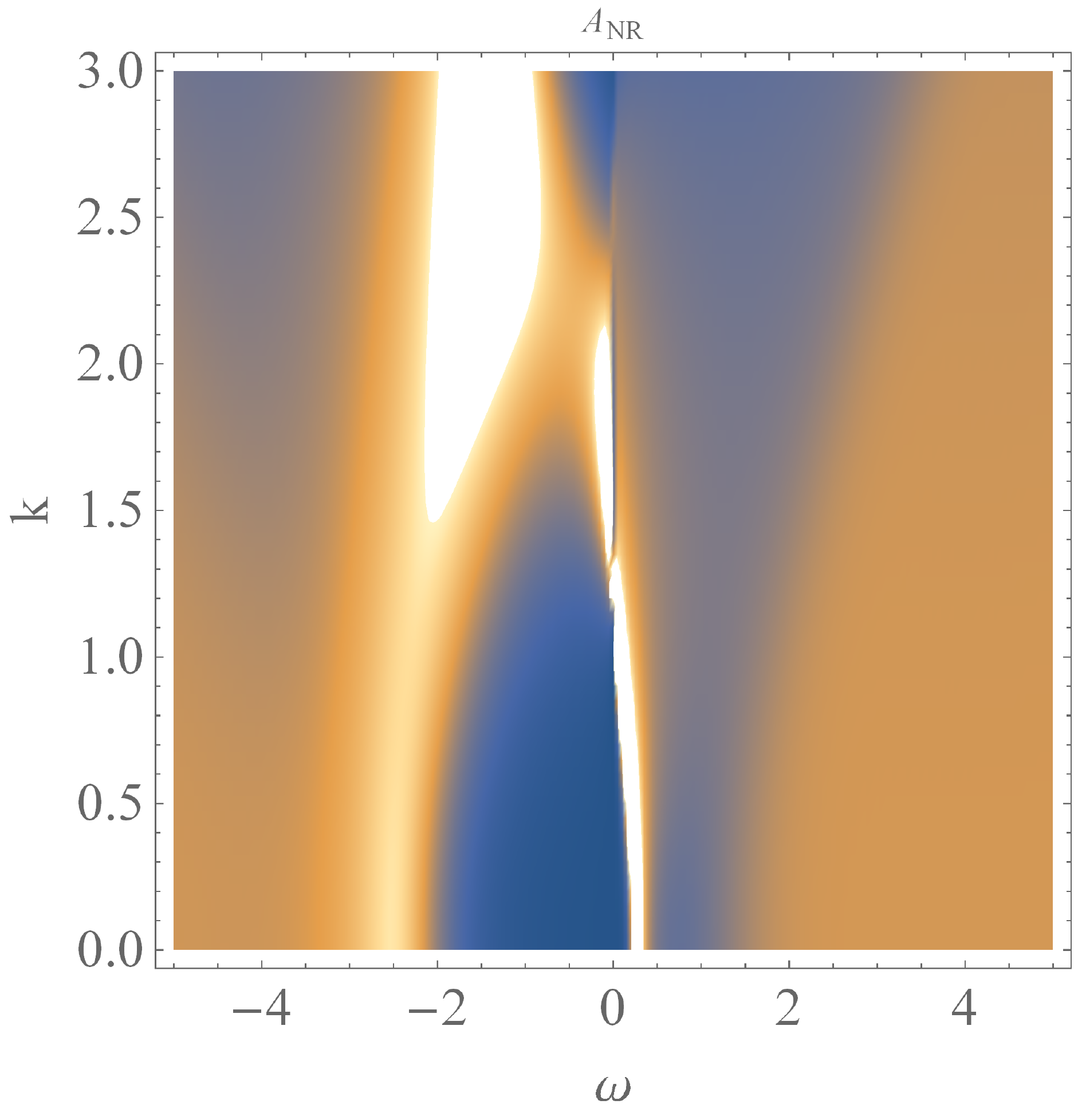}
\ \\
\includegraphics[scale=0.3]{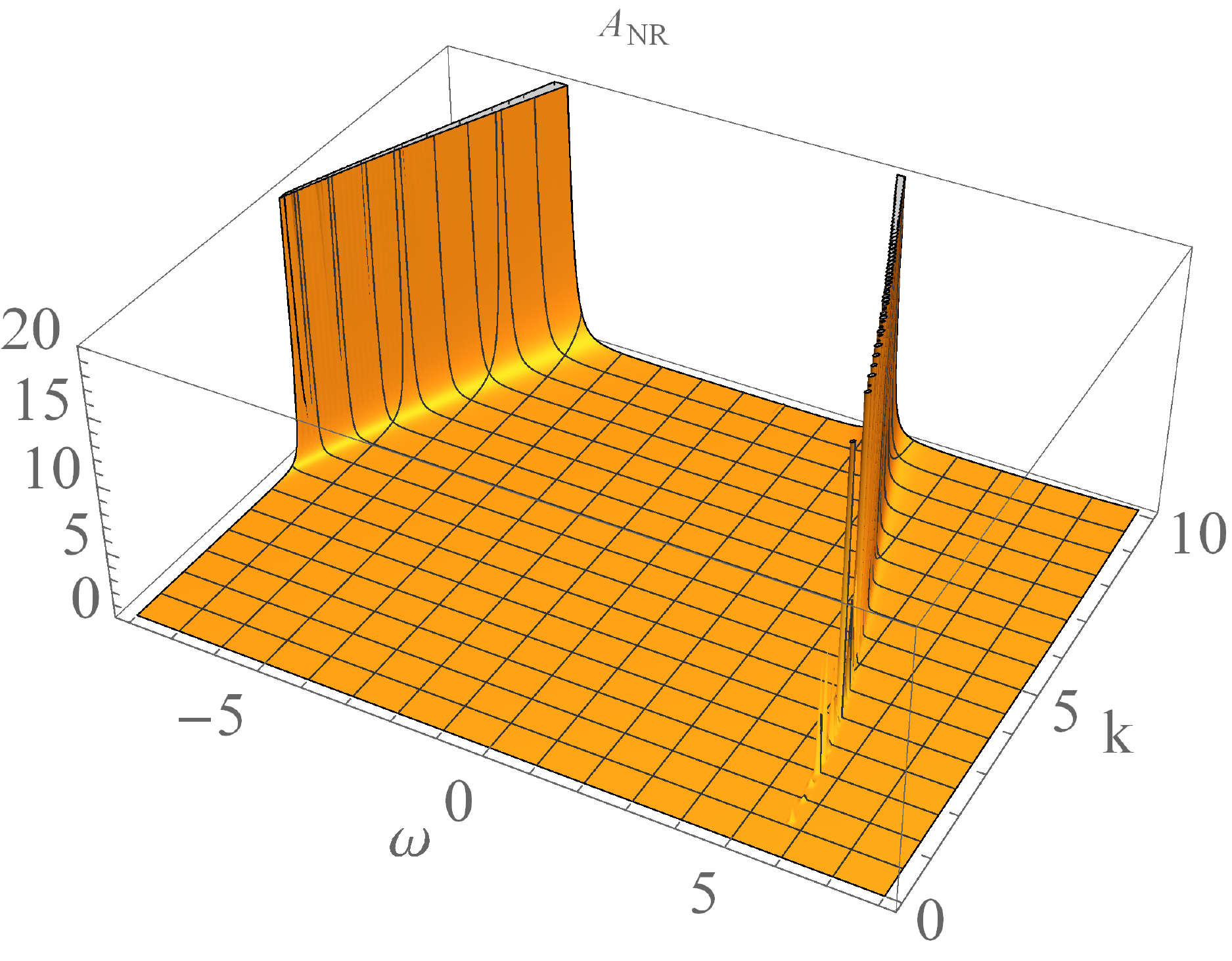}\ \hspace{0.4cm}
\includegraphics[scale=0.26]{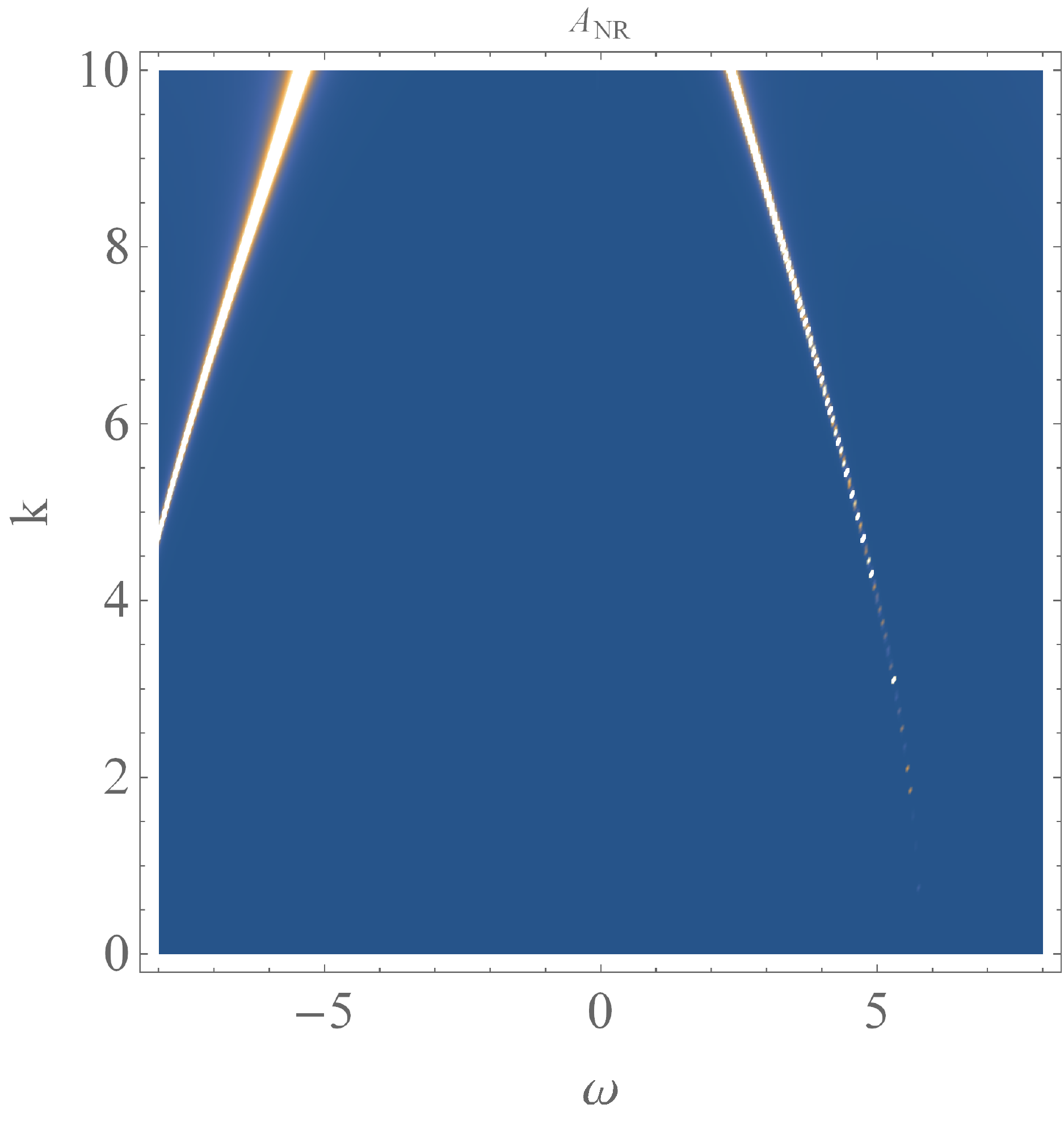}
\ \\
\caption{\label{ANp2p16}
Plots above: The 3d and density plots of the non-relativistic spectral function $A(\omega,k)$ for $p=2$ and $\gamma=0.45$.
Plots below: The 3d and density plots of the non-relativistic spectral function $A(\omega,k)$ for $p=20$ and $\gamma=6$.
All the other parameters are fixed as $q=1$, $m=0$, and $T=0$.}}
\end{figure}

In this section, we shall explore the non-relativistic fermionic spectrum dual to BI-AdS black hole.
FIG.\ref{3DNR} shows the 3d and density plots of the non-relativistic spectral function $A(\omega,k)$ for $p=0$ and $\gamma=0.45$,
in which a holographic flat band emerges as revealed in other geometries \cite{Laia:2011zn,Li:2011nz,Li:2011sh,Alishahiha:2012nm,Li:2012uua,Wu:2013vma}.
The band is mildly dispersive at low momentum region while dispersionless at high momentum region.
Also the flat band is located at $\omega\simeq -1.78$,
which is just the effective chemical potential $\mu_{\gamma}$ (Eq.(\ref{mugamma})).
At the same time, instead of the Fermi surface in the relativistic fermionic spectrum,
only a small bump is developed at the Fermi level in the non-relativistic one.

In \cite{Li:2011nz,Kuang:2012tq}, it has been shown that when the dipole coupling term is turned on,
the flat band is robust and also locates at $\omega\simeq -q\mu$.
In addition, the Fermi surface emerges again as the dipole coupling $p$ increases,
which is different from the relativistic case that a gap forms.
However, the thing becomes subtle when the dipole coupling is turned on in BI-AdS background.
FIG.\ref{ANp2p16} shows that for small $p$ and $\gamma$ (the plot above in FIG.\ref{ANp2p16}),
the Fermi surface sprouts up again as seen in \cite{Li:2011nz,Kuang:2012tq},
while for large $p$ and $\gamma$ (the plot below in FIG.\ref{ANp2p16}),
a gap produces again.
In the following, we shall quantitatively explore the effects from BI term.

\begin{figure}
\center{\includegraphics[scale=0.6]{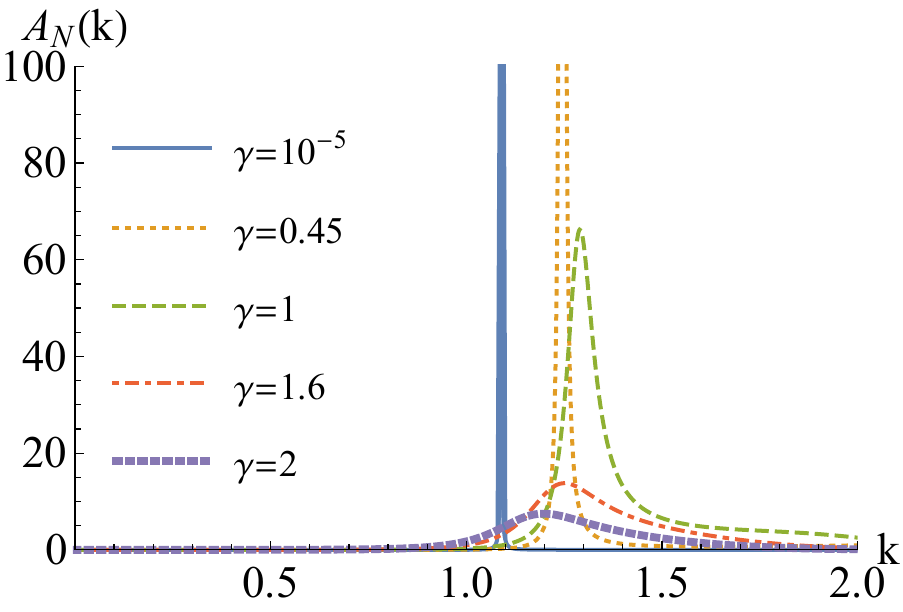}\ \hspace{0.4cm}
\includegraphics[scale=0.4]{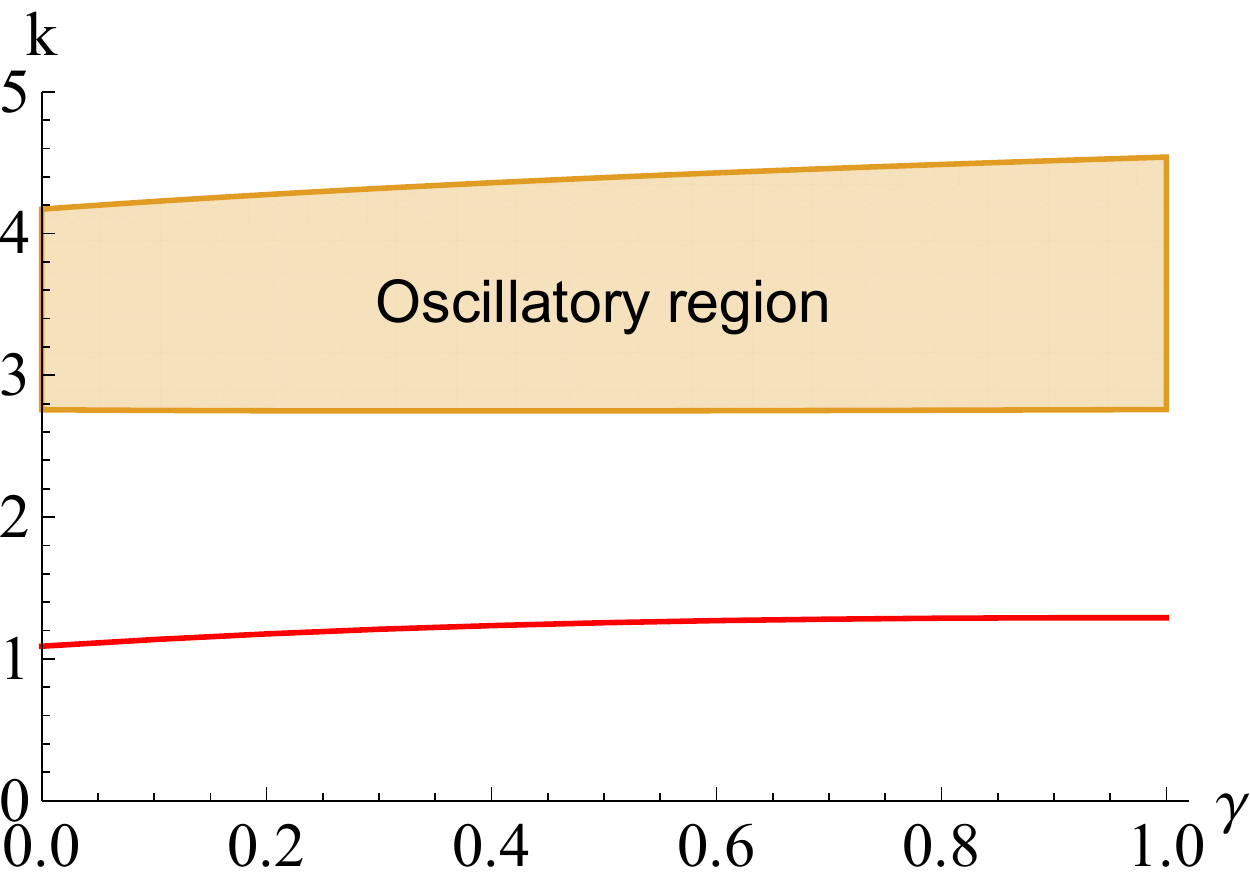}
\ \\
\caption{\label{ANk}
Left plot: The non-relativistic spectral function $A_N(k)$ with $p=2$ at $\omega=-10^{-6}$ for sample BI parameter $\gamma$.
With the increase of $\gamma$, the quasi-particle-like peaks develop into some small bumps.
Here, we have imposed the boundary condition at $u=1-10^{-6}$.
Right plot: The red line is the relation between $\gamma$
and the location of the peak of non-relativistic spectral function $A_N(k)$ ($\omega=-10^{-6}$) for $p=2$.
The yellow zone is the oscillatory region.
The peak don't touch the oscillatory region.
All the other parameters are fixed as $q=1$, $m=0$ and $T=0$.}}
\end{figure}

We firstly fix the dipole coupling $p=2$.
For $\gamma=10^{-5}$, we find that a sharp quasi-particle-like peak emerges at $k_F\simeq 1.0892$ (left plot in FIG.\ref{ANk}),
which is consistent with that found in \cite{Li:2011nz}.
With the increase of $\gamma$, the peak gradually becomes disperse
and finally develops into some small bumps\footnote{Note that for the relativistic fermionic spectrum, we don't observe the dispersive peak before the peak enters into the oscillatory region.
This difference between the relativistic fermionic spectrum and the non-relativistic one calls for further understanding.}.
But even if we further increase $\gamma$,
we cannot see the formation of gap.
Also we show the relation between the BI parameter $\gamma$
and the location of the peak of non-relativistic spectral function $A_N(k)$ ($\omega=-10^{-6}$) in right plot in FIG.\ref{ANk},
in which we can see that the peak doesn't touch the oscillatory region.

\begin{figure}
\center{\includegraphics[scale=0.6]{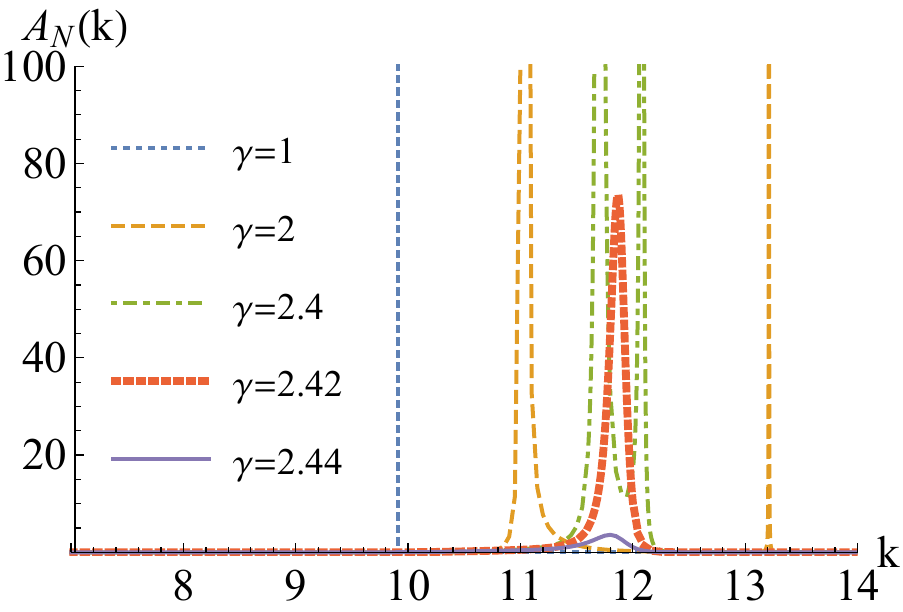}\ \hspace{0.4cm}
\includegraphics[scale=0.4]{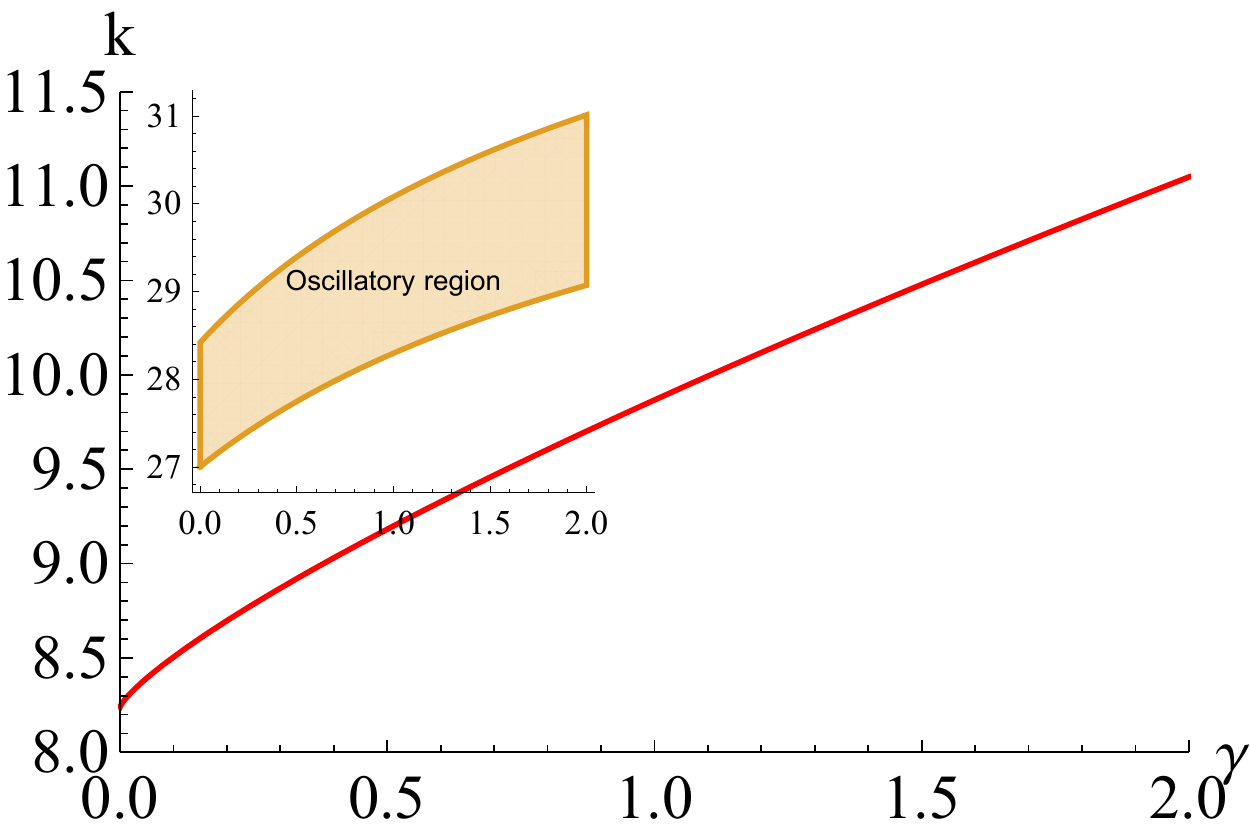}
\ \\
\caption{\label{ANkp16}
Left plot: The non-relativistic spectral function $A_N(k)$ with $p=16$ at $\omega=-10^{-6}$ for sample BI parameter $\gamma$.
Here, we have imposed the boundary condition at $u=1-10^{-6}$.
Right plot: The red line is the relation between the BI parameter $\gamma$
and the Fermi momentum $k_F$ for $p=16$ in the region $\gamma\in[0,2]$.
The inset exhibits the oscillatory region.
The peak doesn't touch the oscillatory region.
All the other parameters are fixed as $q=1$, $m=0$ and $T=0$.}}
\end{figure}

Subsequently we furthermore increase $p$ to see what happens.
The left plot in FIG.\ref{ANkp16} exhibits the non-relativistic spectral function $A_N(k)$ with $p=16$ at $\omega=-10^{-6}$ for sample $\gamma$.
With the augment of $\gamma$, the sharp quasi-particle-like peak also becomes disperse like the case of $p=2$.
But for the case of $p=16$, before developing into small bump, with the increase of $\gamma$, the peak firstly bifurcates and then they again combine into one.
It is a new phenomenon calling for further understanding.
The right plot in FIG.\ref{ANkp16} shows the relation between $\gamma$
and the Fermi momentum $k_F$ for $p=16$ in the region $\gamma\in[0,2]$,
which indicates the Fermi momentum $k_F$ increases as $\gamma$ increases\footnote{We would also like to point out that
when $\gamma$ is beyond some critical value, the location of peak begins to decrease and the peak gradually becomes disperse.}.
In addition, the peak doesn't touch the oscillatory region as that of $p=2$.


\begin{figure}
\center{\includegraphics[scale=0.5]{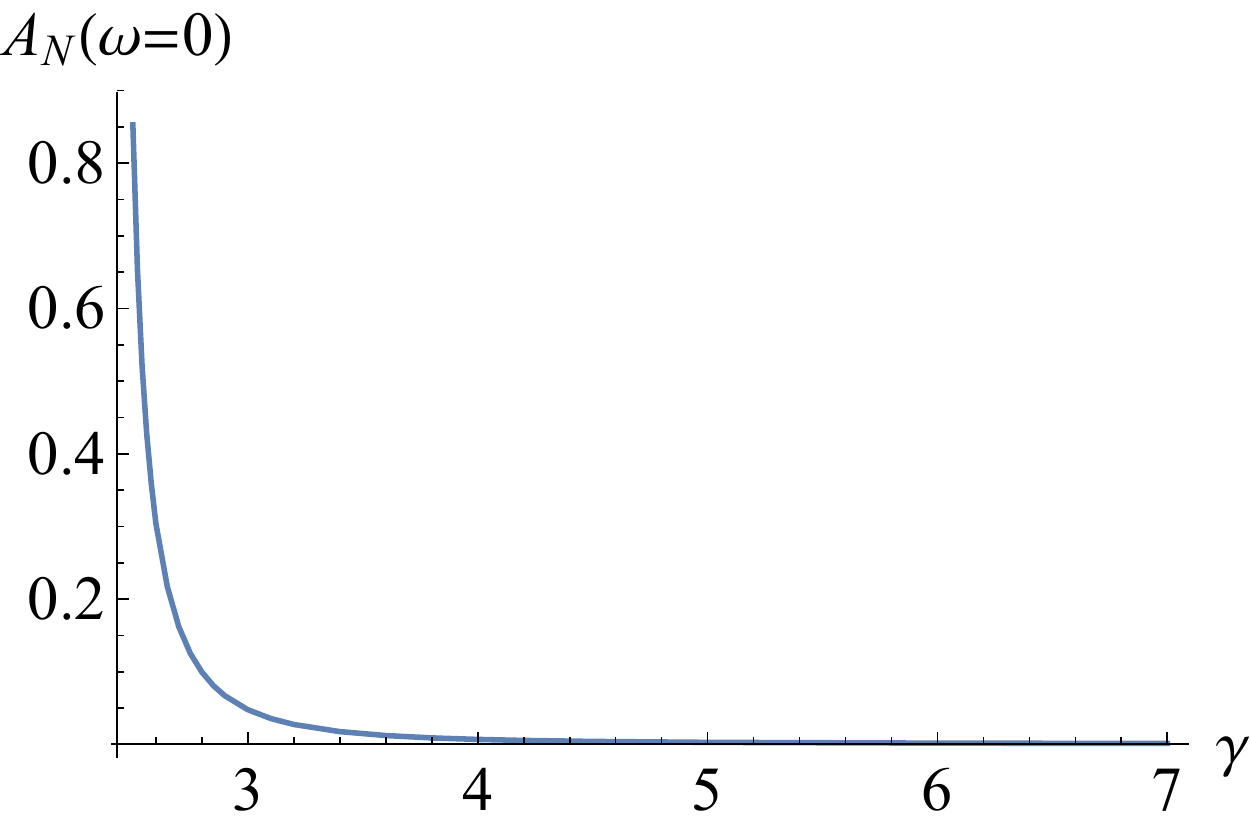}\ \hspace{0.4cm}
\includegraphics[scale=0.5]{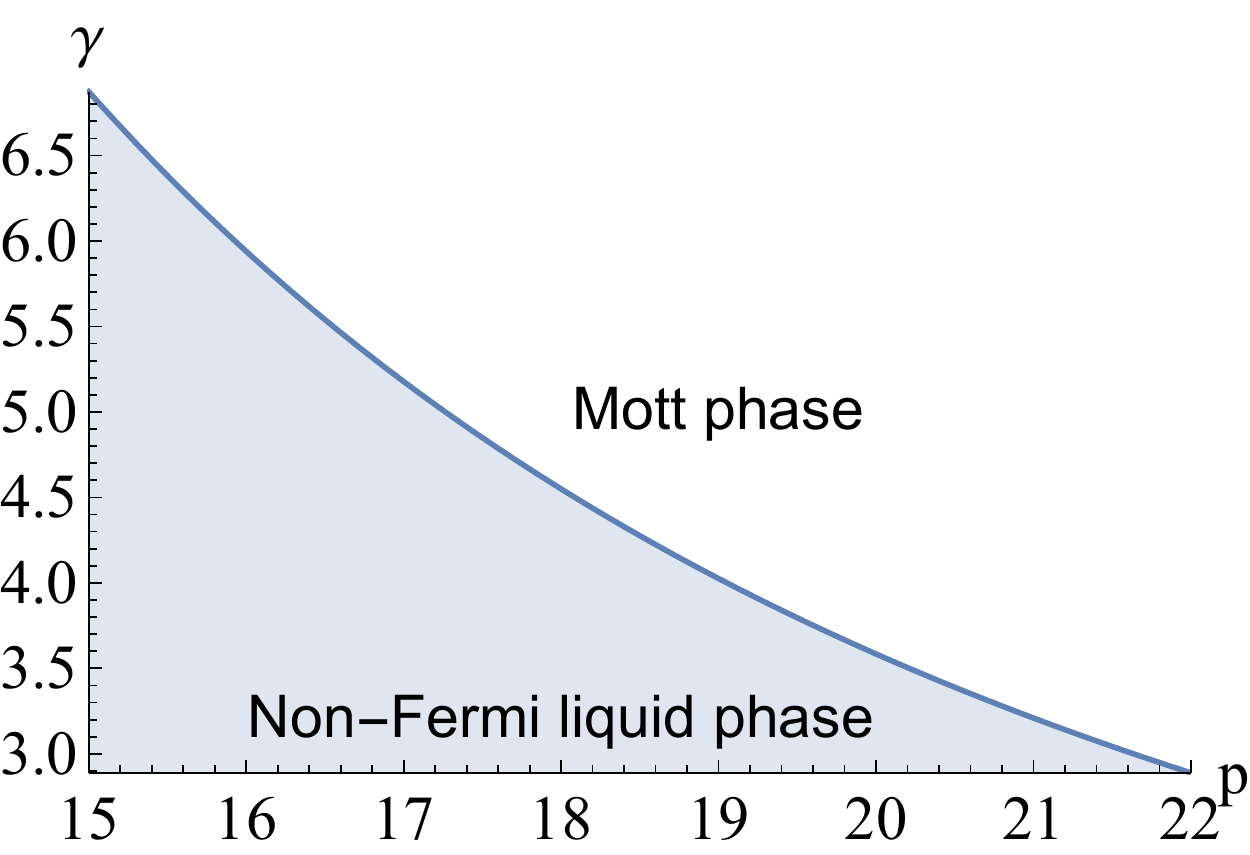}
\ \\
\caption{\label{ANvsgammap16}
Left plot: The relation between DOS of the non-relativistic fermionic spectrum at $\omega=0$ and $\gamma$ for $p=16$.
There is a critical value $\gamma_{c}\simeq 5.6$, above which a gap opens.
Right plot: The phase diagram $(p,\gamma)$ at the region $p\in [15,22]$.
The blue line is the critical line, above which the Mott gap opens.
All the other parameters are fixed as $q=1$, $m=0$ and $T=0$.}}
\end{figure}

Now, we shall turn to explore the formation of Mott gap at the non-relativistic fermionic fixed point in BI-AdS background.
In FIG.\ref{ANp2p16}, it has been revealed that for the large $p$ non-relativistic fermionic system in BI-AdS background,
a gap opens when $\gamma$ is beyond some critical value.
Here, we quantitatively plot the relation between DOS at $\omega=0$ and $\gamma$ for $p=16$ in FIG.\ref{ANvsgammap16},
which shows that the DOS decreases with the increase of $\gamma$ and the critical value of gap formation can be numerically determined
as $\gamma_{c}\simeq 5.6$. Above this critical value, a gap opens.
Furthermore, for $p\in[15,22]$, the phase diagram $(p,\gamma)$ is exhibited in FIG.\ref{ANvsgammap16},
from which we see that for the fixed $p\in[15,22]$ a phase transition from non-Fermi liquid phase to Mott phase as $\gamma$ becomes large.
It indicates that for large $p$ non-relativistic fermionic system, the BI parameter $\gamma$ plays the key role in the formation of gap.

Finally, we would like to point out that it is in the region $p\in[15,22]$ that the Mott gap is observed as $\gamma$ is beyond some critical value.
When $p$ lies outside this region ($p\in[15,22]$), it is hard to determine the critical line between non-Fermi liquid phase and Mott phase
since the numerics become heavier at large $p$ or large $\gamma$.
But we would like to point out that the result presented in the right plot in FIG.\ref{ANvsgammap16} is not
contradict with the previous study in \cite{Li:2011nz,Kuang:2012tq},
in which the they only explore the case of $p\leq 8$.
To address this problem for larger $p$, the analytical exploration needs to be developed as that in \cite{Faulkner:2009wj,Edalati:2010ge,Wu:2013vma}.
We leave this for future study.

\section{Conclusions and discussion}\label{ConD}

In this letter, we systematically explore the fermionic spectrum dual to BI-AdS black hole.
It is the first time that the effects on fermionic spectrum from the corrections of non-linearity of gauge field
are worked out. Our results illuminate some new phenomenon in fermionic spectrum
different from that from GB correction or other geometries \cite{Wu:2011bx,Kuang:2012ud,Wu:2011cy,Li:2011sh,Gursoy:2011gz,Alishahiha:2012nm,Wu:2012fk,Fang:2012pw,Wen:2012ur,Li:2012uua,Kuang:2012tq,Wang:2013tv,
Wu:2013xta,Wu:2013oea,Wu:2014rqa,Wu:2013vma,Fang:2013ixa,Fan:2013zqa,Kuang:2014pna,Kuang:2014yya,Ling:2014bda,Vanacore:2015,Fang:2014jka,Fang:2015vpa,Fang:2015dia}.
We summarize our main findings as follows.
\begin{itemize}
  \item The relativistic fermionic system dual to BI-AdS black hole exhibits non-Fermi liquid behavior.
  The BI parameter aggravates the degree of deviation from Fermi liquid.
  While for the non-relativistic fermionic system, the quasi-particle peak develops into the small bump,
  which is consistent with that found in RN-AdS and dilaton black hole.
  \item For the relativistic fermionic system with dipole coupling, with the
  increase of BI parameter the formation of gap gradually becomes hard.
  It indicates that the BI parameter hinders the formation of Mott gap.
  \item For the non-relativistic fermionic system with large dipole coupling in BI-AdS background,
  with the increase of BI parameter, the gap emerges against, which is a new phenomenon.
\end{itemize}

Still there are a lot directions worthy of further exploration in the future.
Firstly, it is valuable to analytically work out the low frequency behavior
of the non-relativistic fermionic system with dipole coupling
in BI-AdS black hole following that in \cite{Wu:2013vma},
in which the low frequency behavior has been obtained for the non-relativistic fermionic system.
Secondly, it is also interesting to explore the phase diagram $(T,\gamma)$ to further see the role BI parameter playing.
Thirdly, it maybe give more rich physics to explore the fermionic spectrum dual the gravity background with Weyl correction,
which has exhibited the strong to weak coupling transition in the dual field theory \cite{Myers:2010pk,Wu:2010vr,Ma:2011zze}.
The related works in these directions are under progress.
Fourthly, although the fermionic spectrum exhibits the emergence of a gap when the dipole coupling term is introduced,
it is important to note that the electric conductivity is not be gapped, since the underlying AdS$_2$ IR geometry is not a cohesive phase.
In this sense these systems are not real Mott insulator.
To implement a real Mott insulator with gapped electric conductivity and fermionic spectrum, we can introduce the probe fermion
in the gapped geometry for the gauge field, for instance in \cite{Ling:2015epa,Ling:2015exa,Kiritsis:2015oxa} and explore the fermionic excitation.

\begin{acknowledgments}
We are grateful to the anonymous referees for valuable suggestions and comments,
which are important in improving our work.
This work is supported by the Natural
Science Foundation of China under Grant Nos. 11305018, 11275208,
Program for Liaoning Excellent Talents in University (No. LJQ2014123)
and the grant (No.14DZ2260700) from the Opening Project of Shanghai Key Laboratory of High Temperature Superconductors.
\end{acknowledgments}

\end{document}